\documentclass[aps,prl,twocolumn,amsmath,amssymb,reprint,groupedaddress]{revtex4-1}
\pdfoutput=1

\usepackage{float}
\usepackage{graphicx}  
\usepackage{xcolor}
\usepackage{dcolumn}   
\usepackage{bm}        
\usepackage{braket}
\usepackage{amssymb}   
\usepackage{epstopdf}
\usepackage{xfrac}
\usepackage{subfigure}
\usepackage{anyfontsize}

\usepackage{amsmath}
\usepackage{amsfonts}
\usepackage{bbm}
\definecolor{darkblue}{rgb}{0.1,0.2,0.6}
\definecolor{darkred}{rgb}{0.8,0.1,0.2}
\usepackage[colorlinks,citecolor=darkblue,linkcolor=black,urlcolor=darkblue]{hyperref} 
\usepackage{tikz}
\usepackage{enumerate}
\usepackage{setspace}
\usepackage{url}  
\usepackage{mathrsfs}

\newcommand{\Z}{\mathbb{Z}}
\newcommand{\R}{\mathbb{R}}
\newcommand{\Rp}{{\cal R}_\text{part}}

\newcommand{\kv}{\textbf{k}}

\usepackage{pdfpages} 
\makeatletter
\AtBeginDocument{\let\LS@rot\@undefined}
\makeatother

\begin{document}
\title{Many-body topological invariants for fermionic symmetry-protected topological phases}
\author{Hassan Shapourian}
\author{Ken Shiozaki}
\author{Shinsei Ryu}
\affiliation{Department of Physics, University of Illinois at Urbana-Champaign, Urbana Illinois 61801, USA}
\date{\today}
\begin{abstract}
We define and compute many-body topological invariants of 
interacting fermionic symmetry-protected topological phases, 
protected by an orientation-reversing symmetry,
such as time-reversal or reflection symmetry.
The topological invariants are given by partition functions obtained
by a path integral on unoriented spacetime
which, as we show, can be computed for a given ground state wave function 
by considering a non-local operation, ``partial'' reflection or transpose.
As an application of our scheme, we study the $\mathbb{Z}_8$ and $\mathbb{Z}_{16}$ 
classification of topological superconductors in one and three dimensions. 
\end{abstract}
\maketitle

The Thouless-Kohmoto-Nightingale-den Nijs (TKNN) formula~\cite{TKNN,KOHMOTO1985343}
is the prototype for topological characterization of phases of matter.
It relates the quantized Hall conductance to the (first) Chern number defined for Bloch wave functions.
At the level of many-body physics, 
the quantized Hall conductance can also be formulated 
in terms of ground state wave functions in the presence of twisted boundary conditions
(``the many-body Chern number'')
\cite{Niu1985}.
In contrast to local order parameters, the TKNN integer distinguishes
different quantum phases of matter
by focusing on their global topological properties.

More recently, the discovery of topological insulators and superconductors~\cite{KaneRev,ZhangRev} 
has led to a new research frontier, 
generally referred to as symmetry protected topological (SPT) phases. 
These phases are adiabatically connected to 
topologically trivial states,
i.e., atomic insulators which can be represented as simple
product states without any entanglement.
Nevertheless, they are topologically distinct once
a symmetry condition, e.g., time-reversal symmetry, is imposed.
A complete classification of the noninteracting fermionic SPT phases 
protected by non-spatial discrete symmetries
\cite{Schnyder_class,Ryu_class,Kitaev_class}, 
as well as crystalline SPT phases protected by
spatial symmetries~\cite{Fu2011,Chiu2013,Morimoto2013,Shiozaki2014},
were achieved.

However,
it was later discovered that the non-interacting topological classification
is not the full story, and can be dramatically altered
once interaction effects are taken into account~\cite{FidkowskiKitaev1,*FidkowskiKitaev2}. 
Since then, there have been several works which discuss 
the breakdown of the non-interacting classification in the presence of interactions 
\cite{Qi2013, Ryu2012, Gu2014, YaoRyu_int,
Fidkowski2013,Wang1,Wang2, 2014arXiv1406.3032M,
You2014, 2015PhRvB..92h1304I, Hsieh2016, Morimoto2015}.

There are various topological invariants for non-interacting fermionic
SPT phases using single-particle states (e.g., Bloch wave functions). 
For example,
the $\mathbb{Z}_2$-valued topological invariants have been introduced
for topological insulators both in two and three spatial dimensions
\cite{KaneMele1,FuKane,Roy,*Roy2,MooreBalents}.
For topological superconductors protected by time-reversal symmetry,
the integer-valued topological invariants
(``the winding number'') have been introduced
\cite{Schnyder_class}.
However,
the discovered breakdown of non-interacting classification clearly
indicates that the situation at the interacting level is more intricate,
and a general framework to distinguish interacting fermionic SPT phases
is lacking.
This should be contrasted from the quantized Hall conductance,
which can be formulated within many-body physics
without referring to single-particle wave functions
(as it is ultimately related to the response function).

\begin{figure}
\includegraphics[scale=1.1]{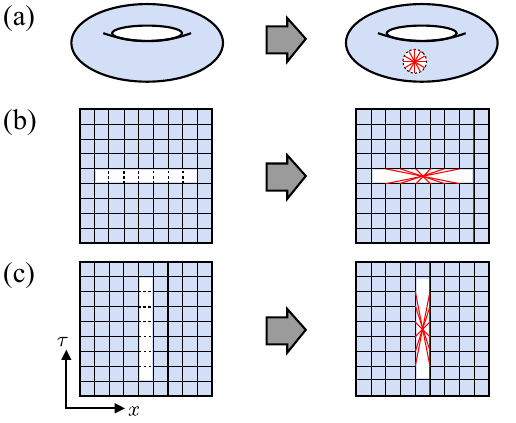}
\caption{\label{fig:crosscap} 
(a) Introducing a cross-cap in the spacetime. 
(b) Partial reflection and (c) partial time inversion.
First and second columns show the original connectivity after cutting and the twisted bonding after gluing, respectively.}
\end{figure}

In this letter, we introduce \emph{many-body} topological invariants
for topological superconductors
protected by an orientation-reversing symmetry,
such as time-reversal or reflection symmetry. 
Our topological invariants do not rely on single-particle descriptions,
and have the same status as the many-body Chern number
\cite{Niu1985}.

The basic strategy behind our construction of many-body topological invariants
can be best illustrated by drawing an analogy with the many-body Chern-number. 
The many-body Chern number is formulated as a response of
the many-body ground state wave functions to the twisted boundary
conditions by $U(1)$ phase.
Here, the $U(1)$ phase is associated with
the symmetry of the system (i.e., the particle number conservation).
Similarly, for phases of matter with more generic symmetry,
one can consider twisting the boundary
condition {\it using the symmetry of the system}.
For SPT phases protected by orientation reversing symmetry,
the symmetry-twisted boundary conditions
naturally give rise to unoriented spacetime manifolds
\cite{2014arXiv1403.1467K,
  *Kapustin2b,*Kapustin2015f,*Kapustin2015b,Freed2014,*Freed2016,
  Hsieh2014,Cho2015,Witten2015, Hsieh2016}.
From the topological quantum field theory description of topological superconductors
\cite{*Kapustin2015f, Witten2015}, 
one expects that the complex phase of
the partition function,
when the system is put on an appropriate unoriented manifold,
is quantized and serves as a topological invariant. 
In the following, we design many-body topological invariants,
such that they return the quantized phase of the partition function.

\emph{One-dimensional topological superconductors
in symmetry classes D+R$_-$ and BDI.--}
We explain our scheme by using the
seminal example of 1D topological superconductors
discussed
by Fidkowski and Kitaev~\cite{FidkowskiKitaev1,*FidkowskiKitaev2}
to show the breakdown of non-interacting
classification (with a slight variation in terms of symmetry requirements):
\begin{align} \label{eq:BdG1d}
\hat{H}= 
-\sum_{x} \Big[t f_{x+1}^\dagger f^{\ }_{x}-\Delta f_{x+1}^\dagger f^\dagger_{x} +\text{H.c.}\Big] -\mu \sum_{x} f_x^\dagger f^{\ }_x,
\end{align}
which describes a superconducting state of spinless fermions.
For simplicity, we take $\Delta$ as a real parameter
and set $\Delta=t$.
The SPT phase in this model, realized when
$|\mu|/t < 2$, is protected either by time reversal
${\cal T} f_x {\cal T}^{-1}= f_{x}$, $\mathcal{T}i\mathcal{T}^{-1}=-i$,
or reflection 
${\cal R} f_x {\cal R}^{-1}= i f_{-x}$.
The former case belongs to symmetry 
class BDI (characterized by time-reversal symmetry where
$\mathcal{T}^2=1$),
while the latter case is referred to as symmetry class ``D+R$_-$''
(class D with reflection symmetry $\mathcal{R}$ satisfying
$\mathcal{R}^2=(-1)^F$ where $F$ is the fermion number).

In the presence of either one of these symmetries,
at the level of non-interacting fermions,
one can introduce the integral topological
index $\nu\in\Z$~\cite{Chiu2013,Morimoto2013,Shiozaki2014}. 
However, the integral classification of the noninteracting fermions collapses 
into the $\Z_8$ classification
in the presence of interactions~\cite{FidkowskiKitaev1,Turner2011}.
Namely,
a stack of eight Majorana chains can be adiabatically
turned into the trivial phase when symmetry preserving interactions are included.

\emph{Spacetime path integral.--} 
As advocated, we now put the system on an unoriented spacetime
and measure the system's response. 
We first present
our many-body invariant using the spacetime path integral and subsequently present the corresponding formula
in the operator formalism, which only involves the
many-body ground states.  

We start from the ordinary Euclidean path integral representation of
the partition function
\begin{align}
&Z=  \text{Tr}(e^{-\beta \hat{H}}) =  \int {\cal D}[\xi] {\cal D} [\bar{\xi}]\ e^{-S[\bar{\xi},\xi]},
\end{align}
where $S[\bar{\xi},\xi]= \int_0^\beta d\tau\ [\bar{\xi} \partial_\tau \xi +
H(\bar{\xi},\xi) ]$,
$\tau$ is the continuous imaginary time variable,
and the Grassmann variables $\xi(\tau,x)$ and $\bar{\xi}(\tau,x)$
are defined at time $\tau$ and real-space position $x$
and obey the anti-periodic temporal boundary condition
$\xi(\tau+\beta) = - \xi(\tau)$,
$\bar{\xi}(\tau+\beta) = - \bar{\xi}(\tau)$. 
The path integral is defined for the spacetime manifold, which is
a torus $T^2$.

In order to create an unoriented spacetime manifold,
we ``modify''
the boundary condition of the path integral, which effectively realizes the real projective plane $\mathbb{R}P^2$. 
The construction of the real projective plane depends crucially
on the type of orientation reversing symmetry
(reflection or time-reversal).
First, for symmetry class D+$R_-$,
we modify the temporal boundary condition at $\tau=\beta\equiv 0$ as 
\begin{align} \label{eq:spacetwist} \xi (\beta+\epsilon,x)&= - \xi (0,x) \to -i {\xi}(0,-x), \nonumber \\
\bar{\xi} (\beta+\epsilon,x)&= -  \bar{\xi} (0,x) \to i \bar{\xi}(0,-x), 
\end{align}
over the interval $|x| < N_{\rm part}/2$ where the reflection
is done with respect to a vertical line crossing the central bond of the segment (Fig.~\ref{fig:crosscap}(b)). Here, $\epsilon$ is the discretization step along the time axis, $\epsilon=\beta/N_t$.
What this procedure does is to first introduce a cut (circle)
in the spacetime path integral  
and then identify opposite points on the circle by
reflection symmetry (Fig.\ \ref{fig:crosscap}(a)).
In short, this ``cut and glue'' process creates
a cross-cap in the spacetime manifold,
which is now topologically equivalent to $\mathbb{R}P^2$.

\begin{figure}
\includegraphics[scale=1.05]{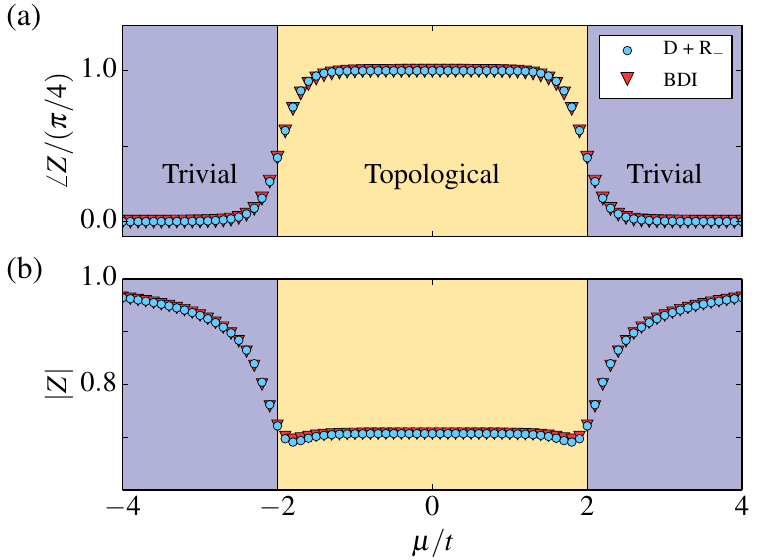}
\caption{\label{fig:spacetime} Phase and amplitude of the partition function in the presence of spatial (Fig.~\ref{fig:crosscap}(b)) and temporal (Fig.~\ref{fig:crosscap}(c)) cross-caps (Details in Appendix B. In short, we write the partition function in terms of a Pfaffian and extract the SPT phase by evaluating the ratio of the two Paffafians in the presence and absence of a cross-cap.)
Here, we set $\beta=10$, $N_t=200$ and $N=40$. We put $N_{t,\text{part}}=100$ and $N_\text{part}=20$ for time inversion (BDI) and spatial reflection (D+R$_{-}$.), respectively. }
\end{figure}

As for symmetry class BDI,
we begin by noting that
in the path-integral formalism,
the time-reversal symmetry which is an antiunitary transformation in the operator formalism,
should be implemented as an invariance under a change of path-integral variables.
For our model,
time-reversal transformation is equivalent to the change of Grassmann fields as in
$
 \xi(\tau,x) \to i\bar{\xi}(\beta-\tau,x)
$,
$
 \bar{\xi}(\tau,x) \to i\xi(\beta-\tau,x).  
$
One can check that this transformation leaves the Hamiltonian 
\eqref{eq:BdG1d} or, in fact, generic bilinear forms 
$H(\bar{\xi},\xi)= \sum_{x,x'} [t_{xx'} \bar{\xi} (x)\xi(x') + \Delta_{xx'} \bar{\xi} (x)\bar{\xi}(x')+\text{H.c.}]$
invariant. 
It is easy to see that possible two-body interaction terms 
such as $(\bar{\xi}\xi)^2$ are also invariant. 
Similar to the cross-cap introduced by twisting the temporal boundary 
by reflection, 
one can twist the spatial boundary condition using
time reversal,
\begin{align} \label{eq:timetwist}
\xi (\tau,N+1)&= -  \xi (\tau,0) \to -i \bar{\xi}(\tilde{\tau},0), \nonumber \\
\bar{\xi} (\tau,N+1)&= -  \bar{\xi} (\tau,0) \to -i {\xi}(\tilde{\tau},0), 
\end{align}
over a time interval $t_1<\tau<t_2$ where $0< t_1,t_2<\beta$ and the time inversion $\tau\to \tilde{\tau}$ is performed with respect to the central line $\tau=(t_1+t_2)/2$ of the interval (Fig.~\ref{fig:crosscap}(c)). 

The topological quantum field theory description of 
topological superconductors implies that
the partition function $Z$ of the canonical model (\ref{eq:BdG1d})
on $\mathbb{R}P^2$ is given by $Z\sim e^{i2\pi/8}$
(the eighth root of unity) in the topological regime, 
corresponding to $\nu=1 \in \Z_8$,
and $Z\sim 1$ in the trivial regime, corresponding to $\nu=0$.
Moreover, the complex phase is additive, i.e., stacking $n$ copies of Majorana
chain (\ref{eq:BdG1d}) results in $Z\sim e^{i2\pi n/8}$,
and for instance, we have $Z\sim 1$ for $n=8$ that is indicative of the trivial
phase.
This is how the $\Z_8$ cyclic group is understood in our scheme.

The numerically computed partition function 
in the presence of a cross-cap is shown in Fig.~\ref{fig:spacetime}. 
The results for  
the symmetry classes BDI and D+R$_-$ match with each other. 
Well inside the non-trivial SPT phase $|\mu|/t<2$
(i.e., when the size of the cross-cap is much bigger than the correlation length),
the phase of the partition function is quantized as
$Z\sim  e^{i\frac{\pi}{4}}/\sqrt{2},$
whereas in the trivial phase 
$Z \leq 1$
with no complex phase and
$Z\to 1$ deep inside the trivial phase (animations and Appendix~J). 
The phase factor $e^{i\frac{\pi}{4}}$ is the $\mathbb{Z}_8$ phase associated with the partition function on $\mathbb{R}P^2$. 
We shall discuss more about the amplitude $|Z|$ momentarily.

\begin{figure}
\includegraphics[scale=.5]{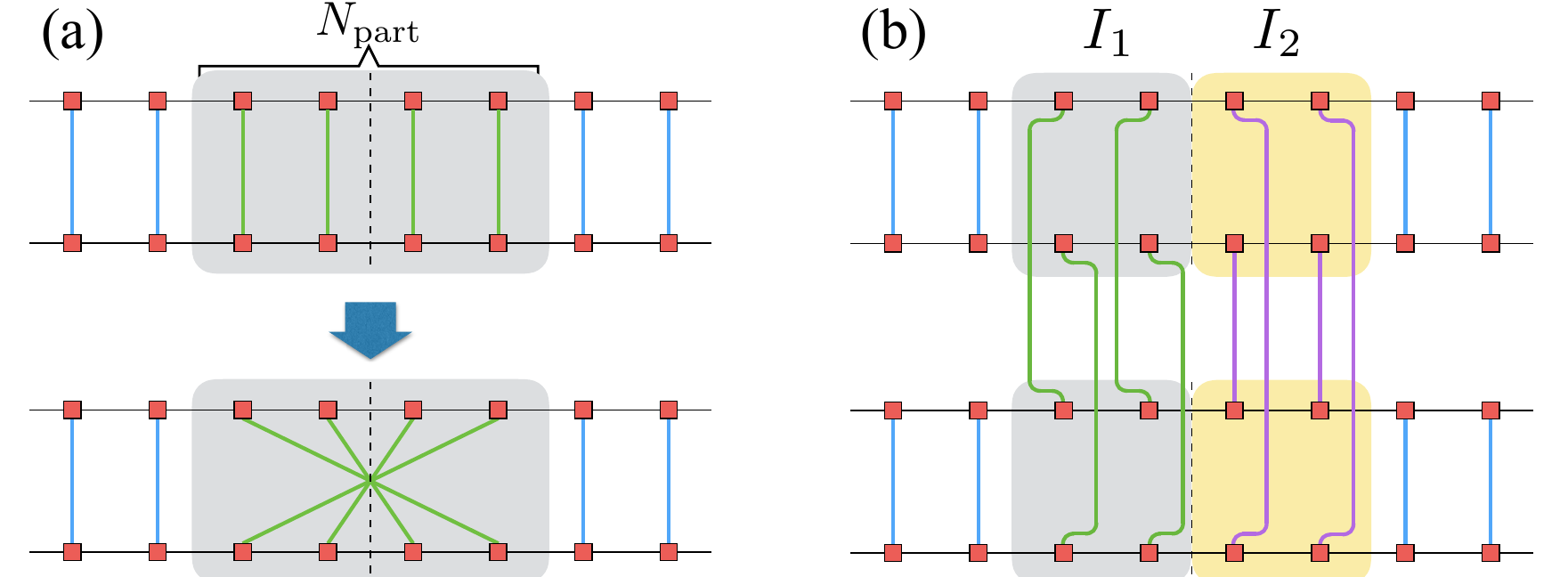}
\caption{\label{fig:MPS_crosscap}
  Schematic representation
  the ground state overlap
  (a) partial reflection $\braket{\Psi|\Rp|\Psi}$,
  (b) partial transpose $\text{tr}\, (\rho_I U_{I_1} \rho_I^{T_1} U_{I_1}^\dagger)$.
  Solid squares represent physical sites and vertical bonds represent how
  physical degrees of freedom are contracted between $\ket{\Psi}$ and $\bra{\Psi}$.}
\end{figure}

\emph{Partial reflection.--}
The cross-cap in the path integral
can be also implemented in terms of  
ground state wave functions of the fermionic SPT phases. 
Let us now discuss this operator formalism. 
The reflection cross-cap can be expressed as 
the expectation value 
of a non-local operator $\Rp$ 
for a given wave function, 
\begin{align} \label{eq:partref}
Z_{\cal{R}}= \bra{\Psi} \Rp \ket{\Psi}
\end{align}
where $\Rp$ is the \emph{partial} reflection operator
which reflects the sites within a segment of
lattice with respect to its central bond
(dashed line in Fig.~\ref{fig:MPS_crosscap} (a)). 
This quantity has been first proposed
as a non-local order parameter to distinguish various topological phases of spin
chains~\cite{Cen2009,Haegeman2012,PTurner_detect1d,Pollmann_protection,Shiozaki-Ryu,stringorder},
The overlap $Z_{\mathcal{R}}$ can be also used as an effective
method to extract the topological 
invariant in the reflection symmetric fermionic SPT phases.

\begin{figure}
\includegraphics[scale=1.05]{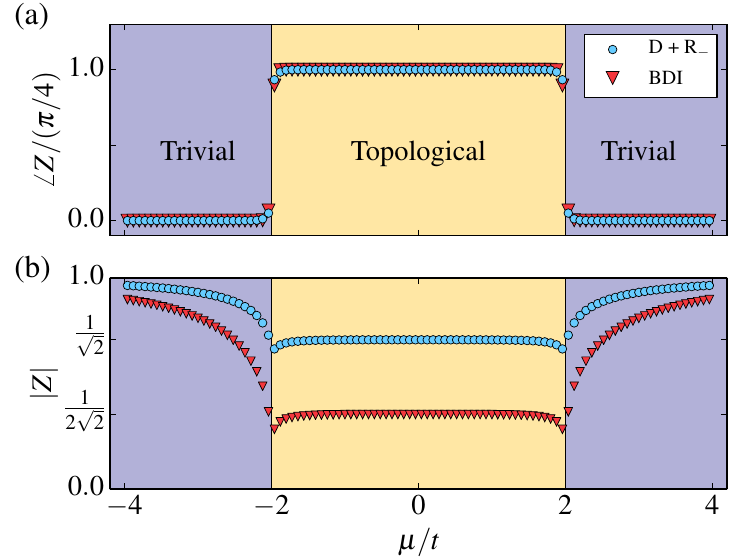}
\caption{\label{fig:Kitaev} Partial reflection (\ref{eq:partref}) for class D+R$_-$ and partial transpose (\ref{eq:ptrans2}) for class BDI in the Kitaev Majorana chain. Here, $N=120$ and $N_\text{part}=60$.} 
\end{figure}

Using the definition of reflection symmetry in the Kitaev chain \eqref{eq:BdG1d},
we can construct $\Rp$ and compute $Z_{\mathcal{R}}$.
The result summarized in Fig.~\ref{fig:Kitaev} confirms that
Eq.~\eqref{eq:partref} shows a similar behavior to its path-integral counterpart.
(An analytical derivation of the same result for the fixed point wave function at $\mu=0$ is provided in Appendix D). More examples including 
the $s$-wave superconducting nanowire construction~\cite{Lutchyn} of Majorana chain,
and the symmetry class A with reflection are provided in Appendix F.

A few remarks regarding the amplitude $|Z|$ are in order. 
First, in the topological phase, the factor $\sqrt{2}$ in the denominator is the quantum dimension of Majorana fermions and physically related to breaking two bonds between the adjacent fermion sites. 
For instance, in the case of class A, 
$|Z|=1/2$ in the topological phase corresponding to 
the quantum dimension of fermionic zero modes (as opposed to Majoranas) at the ends.
Second, in the trivial phase $|Z|$ is $1$ only in the infinite gap limit and the smooth transition from $1/\sqrt{2}$ to $1$ indicates finite size effects.
Another important fact is that $Z$ is a bulk quantity and hence independent of the physical boundary conditions at the ends of a long chain.

\emph{Partial time reversal.--} 
Let us now discuss the way to implement the time-reversal cross-cap in terms of
a given ground state wave function.  
To this end, similar to the partial reflection, we need to introduce partial time-reversal transformation.
Implementation of partial time reversal is slightly more complicated than
partial reflection, since time reversal is anti-unitary
(i.e., it is not clear how to define a partial complex conjugation).
We first explain how to resolve this issue for 1D bosonic SPT phases with time-reversal symmetry ${\cal T}= U {\cal K}$ 
where $\mathcal{K}$ is the complex conjugation and $U$ is a unitary acting on local degrees of freedom. 
Our strategy is to start with the amplitude of full symmetry transformation
$|\braket{\Psi|{\cal T}|\Psi}|$
rather than $\braket{\Psi|{\cal T}|\Psi}$, since the latter is simply not gauge invariant.
Using the definition of time-reversal operator, we write
$|\braket{\Psi|{\cal T}|\Psi}|^2= \text{tr}\, (\rho U \rho^{T} U^\dagger),$
where $\rho=\ket{\Psi}\bra{\Psi}$ is the density matrix and we use the Hermiticity property $\rho^*=\rho^T$.
At this stage, we can conveniently define the topological invariant
in terms of a partial symmetry transformation by introducing the \emph{partial transpose} of the density matrix,
\begin{align} \label{eq:ptrans2}
Z_{\mathcal{T}}=\text{tr}\, \left(\rho^{\ }_I U_{I_1} \rho_I^{T_1} U_{I_1}^\dagger\right).
\end{align}
Here, we consider two adjacent intervals $I_{1,2}$ of the total system $S$,
$\rho_I=\text{tr}_{S\setminus I} (\ket{\Psi}\bra{\Psi})$ is the reduced density matrix for the region $I=I_1\cup I_2$,
and the unitary transformation $U_{I_1}$ acts only in the region $I_1$.
$\rho^{T_1}_I$ is the partial transpose of $\rho_I$,
and for bosonic systems is defined by
\begin{align} \label{eq:ptransb}
\rho_I^{T_1}= \sum_{ijkl} \ket{e_i^1,e_j^2} \braket{e_k^1,e_j^2|\rho_I |e_i^1,e_l^2} \bra{e_k^1,e_l^2}, 
\end{align}
where $\ket{e_j^{1}}$ and $\ket{e_k^{2}}$ denote an orthonormal set of states in
the $I_1$ and $I_2$ regions.
The definition (\ref{eq:ptrans2}) is shown diagrammatically in Fig.~\ref{fig:MPS_crosscap}(b)
and is equivalent to the topological invariant discussed previously in Ref.~\cite{PTurner_detect1d} for spin chains.
Here, it is important that $I_1$ and $I_2$ are adjacent regions;
as we show in Appendix G, 
this configuration is topologically equivalent to introducing a cross-cap in the spacetime
\cite{Shiozaki-Ryu}.

The topological invariant (\ref{eq:ptrans2}) may resemble the \emph{concurrence}
in the context of quantum
information~\cite{Wootters1997,*Wootters1998,Hildebrand,Osborne};
however, Eq.~(\ref{eq:ptrans2}) is different from the concurrence where one
takes the full transpose of the density matrix.
In addition, the eigenvalues of the partially transposed density matrix
$\rho_I^{T_1}$ can be used to define another measure
of quantum entanglement called the \emph{negativity},
which has been shown as an effective probe of the entanglement in mixed states~\cite{Peres1996,Horodecki1996,PlenioEisert1999,Vidal2002,Plenio2005,Calabrese2012}. 

To generalize the expression \eqref{eq:ptrans2} for fermionic systems,  
we need to define a proper partial transpose for fermions. 
This is more transparent when the density matrix is expanded in the coherent state basis
\begin{align} 
\rho_I = \int  d [\bar{\xi},\xi] d [\bar{\chi},\chi]  \ket{\{\xi_j\}} \rho_I \big( \{ \bar{\xi}_j\}; \{ \chi_j \} \big) \bra{ \{ \bar{\chi}_j \} },  \nonumber
\end{align}
where $d [\bar{\xi},\xi] = \prod_j d \bar{\xi}_j d \xi_j e^{- \sum_j \bar \xi_j \xi_j}$ and $\rho_I \big( \{ \bar{\xi}_j\}; \{ \chi_j \} \big) = \Braket{ \{\bar{\xi}_j\} | \rho_I | \{\chi_j\} }$. Using the transformation rules for constructing a time-reversal cross-cap in the path integral formalism (Eq.~(\ref{eq:timetwist})), the analog of Eq.~(\ref{eq:ptransb}) for fermions can be defined as 
\begin{align*} 
U_{I_1} \rho_I^{T_1}  U_{I_1}^\dag &:= \int d [\bar{\xi},\xi] d [\bar{\chi},\chi]   \ket{\{- i \bar{\chi}_j\}_{j \in I_1}, \{ \xi_j\}_{j \in I_2}} \nonumber \\
&\times \rho_I \big( \{ \bar{\xi}_j\}; \{ \chi_j \} \big) 
 \bra{ \{- i \xi_j \}_{j \in I_1}, \{ \bar{\chi}_j \}_{j \in I_2} }.
\end{align*}
The change of variable is effectively equivalent to applying the time-reversal operator only to $I_1$.
An alternative definition of the partial transpose is given in terms of Majorana operators (Appendix G).

Figure~\ref{fig:Kitaev} shows the numerically computed $Z_{\cal T}$ of the Majorana chain (\ref{eq:BdG1d}). 
Well inside the SPT phase, $Z_{\cal T}$ is given by
$Z_{\cal T}\sim e^{i\frac{\pi}{4}}/2\sqrt{2}$
whereas $Z_{\cal T} \leq 1$ in the trivial phase.
Therefore, partial transpose may serve as a non-local operation to obtain the complex phase of the partition function on $\R P^2$ (See Appendix G for analytical derivation and further examples).

\begin{figure}
\includegraphics[scale=1.05]{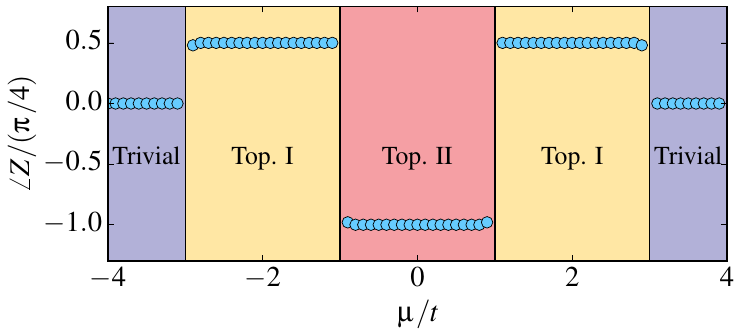}
\caption{\label{fig:3D} 
SPT invariant for the three-dimensional inversion symmetric 
superconductor (class D), Eq.\ \eqref{bdg 3d}.
Top. I (II) corresponds to the phase with odd (even) 
number of gapless Majorana surface states. 
Here, $N=12^3$ and $N_\text{part}=6^3$.}
\end{figure}

\emph{Higher dimensions.--} 
Our scheme, illustrated so far for 1D topological superconductors, can be generalized to 
other fermionic SPT phases, in particular to higher dimensions. 
As an example,
let us consider the inversion symmetric topological superconductor in
class D in 3D (e.g., $^3$He-B phase).
It can be modeled by the BdG Hamiltonian on a cubic lattice, 
which is given in momentum space as
$
\hat H = ({1}/{2}) \sum_\kv \Psi^{\dag}(\kv) h (\kv) \Psi(\kv), 
$
where
$
\Psi^\dag (\kv) = (f_{\uparrow}^\dag(\kv), f_{\downarrow}^\dag(\kv), f_{\downarrow}(-\kv), -f_{\uparrow}(-\kv)), 
$
and
\begin{align}\label{bdg 3d}
h(\kv) 
&= \left[-t( \cos k_x + \cos k_y+ \cos k_z)-\mu\right] \tau_z 
\nonumber \\
& \quad
+ \Delta \left[ \sin k_x \tau_x \sigma_x + \sin k_y \tau_x \sigma_y+ \sin k_z \tau_x \sigma_z\right].
\end{align}
This model is invariant under inversion 
$
{\cal I} f_{\sigma}(x,y,z) {\cal I}^{-1} 
= i f_{\sigma}(-x,-y,-z)
$
(Appendix H).
The topological classification is known to be $\Z_{16}$,
and can be captured by the path integral on the four dimensional real projective
space $\R P^4$,
which can be realized by introducing a cross-cap in $S^4$~\cite{Kirby,Kapustin2015f,Witten2015,Metlitski2015}. 
To define the corresponding topological invariant in terms of the ground state $|\Psi\rangle$, 
we consider the expectation value of the partial inversion
$Z_{\mathcal{I}} = \langle \Psi|\mathcal{I}_{\mathrm{part}}|\Psi\rangle$, which acts on a closed three ball.
The numerically computed SPT invariant is shown in Fig.~\ref{fig:3D}. 
It is quite remarkable that the partial inversion yields the correct $\mathbb{Z}_{16}$ and $\mathbb{Z}_{8}$ complex phases in the topological phases characterized by odd and even number of gapless Majorana surface modes, respectively.

\emph{Discussion.--}
In conclusion, we present an approach to detect interacting SPT phases by 
creating a spacetime cross-cap in the path integral. 
We introduce non-local order parameters partial reflection/transposition
to diagnose many-body SPT phases.
While we use a non-interacting
 fermionic model (the Kitaev chain \eqref{eq:BdG1d}) 
to demonstrate our method, we emphasize that our topological invariants are applicable to interacting models
and can be used in numerical simulations, such as quantum Monte Carlo.
In Appendix F, 
we present the calculation of the topological invariant in 
an interacting Majorana chain, by making use of the known exact expression of the ground state
\cite{Katsura}.
In addition, throughout this letter, we consider BCS mean-field wave functions 
which do not preserve the particle number. 
One important question is whether the partial transformation works 
for particle number conserving systems or not
\cite{Cheng2011,Fidkowski2011,Kraus2013,Ortiz2014,Ortiz2016}. 
As a first step in this direction, we examine the partial reflection
for projected-BCS wave functions,
obtained by projecting the ground state of the mean-field Hamiltonian (\ref{eq:BdG1d}) 
to the space of fixed number of particles.
Using variational Monte Carlo, 
we find that the phase of $Z$ remains quantized as in the mean-field wave function
(Appendix I).
Another important issue is the robustness of SPT invariants in the presence of the random disorder
\cite{ShiozakiHS2016}.


\emph{Acknowledgements.--}
We acknowledge insightful discussions with Takahiro Morimoto, Hosho Katsura and Xueda Wen.
Numerical computation of Pfaffians was carried out by the PFAPACK library~\cite{Wimmer2012}.
This work was supported in part 
by the National Science Foundation Grant No. DMR-1455296,
by the U.S. Department  of  Energy,  Office  of  Science,  
Office  of  Advanced Scientific Computing Research, 
Scientific Discovery  through  Advanced  Computing  (SciDAC)  program
under Award No. FG02-12ER46875,  
and by Alfred P. Sloan Foundation. 
K.S.~is supported by JSPS Postdoctoral Fellowship for Research Abroad.


\bibliography{refs.bib}

\clearpage
 \onecolumngrid


\section*{Supplementary material}

# Supplementary information for "Many-Body Topological Invariants for Fermionic Symmetry-Protected Topological Phases"

Hassan Shapourian, Ken Shiozaki, and Shinsei Ryu

*Department of Physics, University of Illinois at Urbana-Champaign, Urbana Illinois 61801, USA*

The supplementary information is organized as follows: In Appendix A, we explain the basics of the cobordism theory, Appendices B - H are devoted to explaining technical details of construction of SPT invariant from a spacetime cross-cap. In Appendix I, we show that the partial reflection works for the projected BCS wave function onto the sector with a fixed number of fermions. Finally in Appendix J, we explain the details of the attached animations which illustrate the quantization and additivity properties of the proposed topological invariants.

An overview of symmetry classes, spatial dimensions, and SPT invariants introduced in the main text and appendices is outlined in Table I.

| Symmetry class | Spatial dimension | Topological classification | Generating manifold | Spacetime cross-cap |
|---|---|---|---|---|
| D + R$_-$ | 1 | $\Omega_2^{Pin^-} = \mathbb{Z}_8$ | $\mathbb{R}P^2$ | Partial reflection |
| BDI | 1 | $\Omega_2^{Pin^-} = \mathbb{Z}_8$ | $\mathbb{R}P^2$ | Partial time inversion or adjacent partial transpose |
| A + R | 1 | $\Omega_2^{Pin^c} = \mathbb{Z}_4$ | $\mathbb{R}P^2$ | Partial reflection |
| D + I | 3 | $\Omega_4^{Pin^+} = \mathbb{Z}_{16}$ | $\mathbb{R}P^4$ | Partial inversion |

TABLE I. Fermionic symmetry-protected topological phases discussed in the text. For the symmetry classes "D+R$_-$" and "A+R", we follow the notation in Refs. [1–3]. Symmetry class "D+I" means symmetry class D in the presence of inversion symmetry.

## CONTENTS

**Appendix A: Classification of SPT phases and the cobordism theory**

In this appendix, we briefly discuss the relation between the cobordism theory and the classification of SPT phases. As mentioned in the main text, we are interested in the topological phases protected by various symmetry groups, such as time-reversal, charge conjugation, and/or space group symmetry.

The basic principle is that the low-energy long-wavelength physics of the gapped phases of matter can be captured by topological quantum field theories (TQFTs) of some sort. To be more specific, let us consider topological phases protected by a set of global symmetries, called symmetry group $\widetilde{G}$. It is convenient to decompose the symmetry group $\widetilde{G}$ into the part which includes the unitary on-site (or "internal") symmetries ($= G$), and the part which contains the orientation-reversing symmetry transformations over the spacetime manifold, such as time-reversal or parity (other types of spatial symmetries such as point group symmetries were discussed in Ref. [4]). If the system under study contains fermions, the fermion number parity will be included in the former part.

As usual in computing topological invariants of the gapped quantum phases, we couple the matter fields to a background $G$ gauge field. Consequently, one can integrate out the matter fields $\{\phi_i\}$ and derive the effective theory

$$Z(X, \eta, A) = \int \prod_i D\phi_i e^{-S_X(\{\phi_i\}, \eta, A)}, \tag{A1}$$

in terms of the $(d+1)$-dimensional closed spacetime manifold $X$, the background $G$ gauge field $A$ which is coupled to the matter fields. Additionally, the "structure" $\eta$ is also endowed to the manifold, which describes an orientation, a Spin or Spin$^c$ structure for real or complex fermions, respectively. This means that the symmetry group $\widetilde{G}$ of a given gapped (topological) phase enters the effective partition function as the input data $(X, \eta, A)$. Let us remind the reader that for an orientable $(d+1)$-dimensional manifold $X$, the structure group $SO(d+1)$ on its tangent spaces acts on frame fields (vielbein). For a relativistic fermion field on $X$, $SO(d+1)$ is lifted to $Spin(d+1)$. The choice of signs that arises in this lifting defines a *spin structure*. An important remark here is that fermions in conventional condensed matter systems are not always sensitive to spin structures; while in the case of topological phases, where the effective theory is described by relativistic fermions, they do depend on the choice of spin structure. In other words, spin TQFT may emerge in short-range entangled SPT states.

The partition function $Z(X, \eta, A)$, or equivalently the effective action $S_{eff}(X, \eta, A) = -\ln Z(X, \eta, A)$, contains information about topologically non-trivial gapped phases in terms of a topological term, a $U(1)$ phase of the partition function (the imaginary part of the Euclidean effective action), which is insensitive to the metric on $X$ (i.e., it is invariant under diffeomorphisms) as well as small variations of the background gauge field. In fact, the partition function is purely a topological term in the zero correlation length limit. In the case of finite energy gap, the resulting partition functions are expected to behave the same way as long as the correlation length of the system is much shorter than the system size.

In a more mathematical language, we say that the partition function $Z(X, \eta, A)$ defines a TQFT, or more precisely, the so-called $G$-equivariant spin TQFT [5–9]. For a given closed manifold $X$ with background structures specified by $(\eta, A)$, the TQFT returns a topological invariant. In the case of SPT phases (we are dealing with here), there is a unique ground state and the corresponding TQFT is called invertible. Recently, it was proposed that the partition function of such phases depends only on the certain equivalence classes of the background manifolds (with structure $\eta$ and the background gauge field), the cobordism class of $(X, \eta, A)$ [10]. Let us explain the essential concept "cobordism" in this proposal. The cobordism theory is a way to define equivalence classes of $(d+1)$-dimensional manifolds with a given structure and background gauge field configurations: Two $(d+1)$-dimensional spacetime manifolds $X_{1,2}$ (with structure $\eta_{1,2}$ and background fields $A_{1,2}$) are called cobordant when one can find a $(d+2)$-dimensional manifold $Y$ with an appropriate background gauge field that can interpolate $(X_1, \eta_1, A_1)$ and $(X_2, \eta_2, A_2)$, i.e., $\partial Y = X_1 \sqcup X_2$. This relation can then be used to define the equivalence classes. The claim is that the partition function is invariant under cobordism. We illustrate the geometric significance of the cobordism through an example in Fig. 1, which shows the equivalence relation between a single circle and a pair of disjoint circles.

An Abelian group structure can be introduced to the equivalence classes of $(X, \eta, A)$ by taking the disjoint union as an operation. The resulting group is called the cobordism group and denoted by $\Omega^{\text{str}}_{d+1}(BG)$, which is Abelian (e.g., $\mathbb{Z}_n$, $\mathbb{Z}_n \oplus \mathbb{Z}_m$.) Here, $BG$ is the classifying space of $G$ [11, 12]. When there is no symmetry, we simply put a single point as $BG$, $BG = pt$.

In the case of topological theories where the partition function is a pure phase, $Z(X, \eta, A)$ can be viewed as a homomorphism

$$Z: \Omega^{\text{str}}_{d+1}(BG) \to U(1), \quad (X, \eta, A) \mapsto Z(X, \eta, A). \tag{A2}$$

Hence, this fact can be used as a guiding principle to classify possible topological $U(1)$ phases of $Z(X, \eta, A)$ in terms

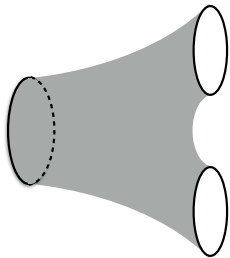

FIG. 1. A cobordism in $(0+1)$-dimensions between a single circle (left) and a pair of disjoint circles (right).

of the cobordism group classification of manifolds with structures [10, 13, 14]. The proposed statement is that the torsion part of the cobordism group Tor $\Omega^{\text{str}}_{d+1}(BG)$ provides a possible classification of topological phases of matter.

According to the above proposal, one realizes that the topological $U(1)$ phase of the path integral evaluated on a suitable manifold with structures $(X, \eta, A)$ can be thought of as a meaningful topological invariant (SPT invariant). In this paper, we prescribe a systematic way to construct the generator $(X, \eta, A)$ of the cobordism group $\Omega^{\text{str}}_{d+1}(BG)$, which we simply call the *generating manifold* here. It is on this manifold that the complex phase associated with the partition function is a "least possible" or "most fundamental" topological $U(1)$ phase. The topological $U(1)$ phases for other possible manifolds $(X, \eta, A)$ are given as an integer multiple of the topological $U(1)$ phase for the generating manifold. A list of generating manifolds for various symmetry classes is provided in Table. I

### Appendix B: Calculation of the partition function on discretized spacetime

In this appendix, we discuss how the computation of the partition function on a discretized spacetime has been done. We also show the transformation rules of Grassmann variables which are equivalent to the action of the time-reversal and reflection symmetries. For numerical purposes, we define the partition function on a spacetime lattice as in

$$Z = \int \prod_{j=1}^{N_t} d[\xi_j] d[\bar{\xi}_j] e^{\sum_j [\bar{\xi}_{j+1} \cdot \xi_j - \bar{\xi}_j \cdot \xi_j]} e^{-\epsilon \sum_j H(\bar{\xi}_{j+1}, \xi_j)}, \tag{B1}$$

where we consider a discretized time axis, $\epsilon = \beta/N_t$, and a real-space lattice where the Grassmann variables $\xi_j(x)$ and $\bar{\xi}_j(x)$ are defined on $j$th time slice and position $x$ on the lattice. We use the shorthand notation $\bar{\xi}_j \cdot \xi_{j'} = \sum_x \bar{\xi}_j(x) \xi_{j'}(x)$. The Hamiltonian operator in this representaion becomes a polynomial, given by

$$e^{-\epsilon H(\bar{\xi}_{j+1}, \xi_j)} = e^{-\bar{\xi}_{j+1} \xi_j} \langle \xi_{j+1} | e^{-\epsilon \hat{H}} | \xi_j \rangle. \tag{B2}$$

The discretized partition function at zero temperature corresponds to the limits $\epsilon \to 0$ and $N_t \to \infty$ such that $\beta = N_t \epsilon \to \infty$.

The partition function can be computed exactly for quadratic Hamiltonians. Let us consider the Kitaev Majorana chain,

$$H(\bar{\xi}_{j+1}, \xi_j) = \sum_x (\boldsymbol{\xi}_j^T(x+1) A \boldsymbol{\xi}_j(x) + \boldsymbol{\xi}_j^T(x) B \boldsymbol{\xi}_j(x)) \tag{B3}$$

where $\boldsymbol{\xi}_j^T(x) = (\bar{\xi}_{j+1}(x), \xi_j(x))$ is a two-component vector (where we use the new notation $\xi_j^1 = \bar{\xi}_{j+1}$ and $\xi_j^2 = \xi_j$) and we introduce the following matrices

$$A = \begin{pmatrix} \Delta & -t \\ t & -\Delta^* \end{pmatrix} \tag{B4}$$

$$B = \frac{1}{2} \begin{pmatrix} 0 & -\mu \\ \mu & 0 \end{pmatrix}. \tag{B5}$$

The full partition function becomes

$$Z = \int \prod_{j=1}^{N_t} d[\xi_j] d[\bar{\xi}_j] e^{\sum_{j,x} \boldsymbol{\xi}_j^T(x)[\frac{i}{2}\sigma_2 - \epsilon B]\boldsymbol{\xi}_j(x)} e^{-\epsilon \sum_{j,x} \boldsymbol{\xi}_j^T(x+1) A \boldsymbol{\xi}_j(x)} e^{\sum_{j,x} \boldsymbol{\xi}_{j+1}^T(x) A_t \boldsymbol{\xi}_j(x)} \quad (B6)$$

where $\sigma_2$ is the second Pauli matrix and

$$A_t = \begin{pmatrix} 0 & 0 \\ 1 & 0 \end{pmatrix}. \quad (B7)$$

The last term $\boldsymbol{\xi}_{j+1}^T(x) A_t \boldsymbol{\xi}_j(x) = \xi_{j+1}^2(x)\xi_j^1(x)$ is the hopping term along the time direction. Time inversion changes the time index $j \to N_t - j$. So, for the last term to be time-reversal invariant it must transforms as in

$$\xi_{j+1}^2(x)\xi_j^1(x) \to i^2 \xi_{N_t-j-1}^1(x)\xi_{N_t-j}^2(x) \quad (B8)$$

which implies the transformation rules

$$\xi_j^1(x) \to i\xi_{N_t-j}^2(x) \quad (B9)$$
$$\xi_j^2(x) \to i\xi_{N_t-j}^1(x) \quad (B10)$$

or in the matrix form

$$\boldsymbol{\xi}_j(x) \to i\sigma_1 \boldsymbol{\xi}_{N_t-j}(x). \quad (B11)$$

Having this transformation rule, let us check whether other terms in action are invariant.

The real space hopping matrix element can be written as

$$A = -it\sigma_2 + \text{Re}[\Delta]\sigma_3 + i\text{Im}[\Delta]\sigma_0. \quad (B12)$$

The Hamiltonian term is always invariant due to anticommuting property of Grassmann variables. Therefore, one can impose twisted boundary conditions across fixed position contours (vertical lines) as introduced in the main text.

The reflection symmetry can be similarly imposed on the action in Eq. (B6),

$$\boldsymbol{\xi}_j(x) \to i\sigma_3 \boldsymbol{\xi}_j(-x). \quad (B13)$$

Using this definition, one can impose twisted boundary condition across fixed time contours (horizontal lines) in the spacetime.

In either cases, the partition function in Eq. (B6) can be computed exactly in terms of a Pfaffian of discretized action matrix, since there are only on-site and hopping terms in both time and space directions.

### Appendix C: BdG Hamiltonian and numerical details

In this appendix, we explain how the BCS wave functions are constructed and how the overlaps are computed. In general, the BdG Hamiltonian in the Nambu basis over real space can be written as

$$\hat{H} = \frac{1}{2} \sum_{i,j} \psi_i^\dagger \begin{pmatrix} t_{ij} & \Delta_{ij} \\ -\Delta_{ij}^* & -t_{ij}^T \end{pmatrix} \psi_j. \quad (C1)$$

where the Nambu spinor is $\psi_i^\dagger = (f_i^\dagger, f_i)$, $t_{ij}^\dagger = t_{ij}$ is a Hermitian matrix and $\Delta_{ij}^\dagger = -\Delta_{ij}^*$ is a skew symmetric matrix. We diagonalize the Hamiltonian to find the eigenvalues $\epsilon_n$

$$\begin{pmatrix} t_{ij} & \Delta_{ij} \\ -\Delta_{ij}^* & -t_{ij}^T \end{pmatrix} \begin{pmatrix} u_{jn} \\ v_{jn} \end{pmatrix} = \epsilon_n \begin{pmatrix} u_{in} \\ v_{in} \end{pmatrix}, \quad (C2)$$

where $n$ is energy eigenvalue index and $i, j$ refer to the sites on the lattice. Taking the complex conjugate of the above expression and exchange the rows yield

$$\begin{pmatrix} t_{ij} & \Delta_{ij} \\ -\Delta_{ij}^* & -t_{ij}^T \end{pmatrix} \begin{pmatrix} v_{jn}^* \\ u_{jn}^* \end{pmatrix} = -\epsilon_n \begin{pmatrix} v_{jn}^* \\ u_{jn}^* \end{pmatrix}. \tag{C3}$$

We define the Bogoliubov operators

$$\begin{pmatrix} b_n \\ b_n^\dagger \end{pmatrix} = \begin{pmatrix} u_{in}^* & v_{in}^* \\ u_{in} & v_{in} \end{pmatrix} \begin{pmatrix} f_i \\ f_i^\dagger \end{pmatrix} \tag{C4}$$

where sum over $i$ is assumed and the inverse relation

$$\begin{pmatrix} f_i \\ f_i^\dagger \end{pmatrix} = \begin{pmatrix} u_{in} & v_{in}^* \\ v_{in} & u_{in}^* \end{pmatrix} \begin{pmatrix} b_n \\ b_n^\dagger \end{pmatrix} \tag{C5}$$

here, sum over $n$ is assumed.

The Hamiltonian in this new basis reads

$$H = \sum_n \epsilon_n (b_n^\dagger b_n - \frac{1}{2}). \tag{C6}$$

Assuming all $\epsilon_n > 0$, the ground state wave function must obey the condition $b_n |\Psi\rangle = 0$. This means that

$$(u_{in}^* f_i + v_{in}^* f_i^\dagger) |\Psi\rangle = 0, \tag{C7}$$

where sum over repeated index $i$ is assumed. The ground state wave function can then be written as

$$|\Psi\rangle = \exp\left(\frac{1}{2} \sum_{i,j} \phi_{ij} f_i^\dagger f_j^\dagger\right) |0\rangle, \tag{C8}$$

where

$$\phi_{ij} = v_{in}^* [u^*]_{nj}^{-1}. \tag{C9}$$

Let us check that this wave function satisfies Eq. (C7).

$$(u_{in}^* f_i + v_{in}^* f_i^\dagger) |\Psi\rangle = (u_{in}^* f_i + v_{in}^* f_i^\dagger) \exp\left(\frac{1}{2} \sum_{i,j} \phi_{ij} f_i^\dagger f_j^\dagger\right) |0\rangle$$

$$= \exp\left(\frac{1}{2} \sum_{i,j} \phi_{ij} f_i^\dagger f_j^\dagger\right) (u_{in}^*(f_i - \phi_{li} f_l^\dagger) + v_{in}^* f_i^\dagger) |0\rangle$$

$$= \exp\left(\frac{1}{2} \sum_{i,j} \phi_{ij} f_i^\dagger f_j^\dagger\right) (u_{in}^* f_i - u_{in}^* [u^*]_{mi}^{-1} v_{lm}^* f_l^\dagger + v_{in}^* f_i^\dagger) |0\rangle$$

$$= 0, \tag{C10}$$

where we use the identity

$$\exp\left(-\frac{1}{2} \sum_{i,j} \phi_{ij} f_i^\dagger f_j^\dagger\right) f_i \exp\left(\frac{1}{2} \sum_{i,j} \phi_{ij} f_i^\dagger f_j^\dagger\right) = f_i - \phi_{li} f_l^\dagger. \tag{C11}$$

One difficulty with the wave function in Eq. (C8) is that if pairing vanishes at any point in the Brillouin zone $u_{jn}$ becomes singular and not invertible. These are simply due to fully occupied states which cannot be represented as a particle-hole mixture (BCS-type wave function).

The inner product of two wave functions in the form of Eq. (C8) is

$$\langle \Psi_1 | \Psi_2 \rangle = \left[ \det(\mathbf{1} + \phi_1^\dagger \cdot \phi_2) \right]^{1/2}, \tag{C12}$$

which can also be written as

$$\langle \Psi_1 | \Psi_2 \rangle = \det \left[ \begin{pmatrix} u_1^* \\ v_1^* \end{pmatrix}^\dagger \cdot \begin{pmatrix} u_2^* \\ v_2^* \end{pmatrix} \right]^{1/2}. \tag{C13}$$

The second expression does not involve inverting $v_i$ matrices and hence more efficient.

It is important to note that the definition in Eq. (C8) can be used to write wave function with a fixed number of particles. We apply a projection operator $\mathcal{P}_n$ to the BCS wave function in Eq. (C8)

$$|\Psi_n\rangle = \mathcal{P}_n |\Psi\rangle, \tag{C14}$$

to project out all other particle numbers. This implies that for a particle number sector with $n$ particles at positions $\{r_1, \ldots, r_n\}$ the wave function can be written as

$$\Psi_n(r_1, \ldots, r_n) = \langle r_1, \ldots, r_n | \mathcal{P}_n | \Psi \rangle$$
$$= \text{Pf}[\phi_{ij}(r_1, \ldots, r_n)] \tag{C15}$$

where the indices $i$ and $j$ refer to the real-space positions, i.e., the corresponding rows and columns of $\phi_{ij}$, i.e. $i, j \in \{r_1, \ldots, r_n\}$, must be chosen.

### Appendix D: Partial reflection of fixed-point wave function

In this appendix, we write down the explicit form of the fixed-point Majorana chain Hamiltonian and its ground state wave function and analytically derive the partial reflection in this case. In the Majorana basis, the Majorana chain Hamiltonian (Eq. (1) of main text) takes the form

$$\hat{H} = \frac{i}{2} \sum_{x=1}^{N} [(\Delta + t) c_x^R c_{x+1}^L + (\Delta - t) c_x^L c_{x+1}^R] - \frac{\mu}{2} \sum_x (1 + i c_x^L c_x^R), \tag{D1}$$

which is derived by the following substitutions in the original Hamiltonian,

$$f_x^\dagger = \frac{e^{-i\phi/2}}{2}(c_x^L - i c_x^R), \quad f_x = \frac{e^{i\phi/2}}{2}(c_x^L + i c_x^R). \tag{D2}$$

From now on, we take $\Delta$ to be real ($\phi = 0$) for simplicity, while the result does not depend on this choice of the gauge. The fixed-point limit (zero correlation length) is defined when $\Delta = t$ and $\mu = 0$. In this limit, the Hamiltonian reads

$$H = i \sum_{x=1}^{N-1} c_x^R c_{x+1}^L \tag{D3}$$

$$= \sum_{x=1}^{N-1} (2 g_x^\dagger g_x - 1) \tag{D4}$$

where the new fermion operators are

$$g_x = \frac{1}{2}(c_x^R + i c_{x+1}^L) = \frac{i}{2}(f_x^\dagger - f_x + f_{x+1}^\dagger + f_{x+1}), \tag{D5}$$

$$g_x^\dagger = \frac{1}{2}(c_x^R - i c_{x+1}^L) = \frac{i}{2}(f_x^\dagger - f_x - f_{x+1}^\dagger - f_{x+1}), \tag{D6}$$

and their inverses are

$$c_x^R = g_x + g_x^\dagger, \quad c_{x+1}^L = i(g_x^\dagger - g_x). \tag{D7}$$

The ground state can be found easily using the Hamiltonian in Eq. (D4), by requiring that $g_x |\Psi\rangle = 0$ which is the vacuum of $g_x$ operators. Therefore, the wave function can be written in terms of

$$\begin{aligned} |\Psi_\pm\rangle &= \prod_{x=1}^{N}(1 \pm f_x^\dagger)|0\rangle \\ &= (1 \pm f_1^\dagger)(1 \pm f_2^\dagger)\dots(1 \pm f_N^\dagger)|0\rangle, \end{aligned} \tag{D8}$$

where we assume the convention in which the fermion index is in ascending order. It is easy to check that $g_x |\Psi_\pm\rangle = 0$ for $1 \leq x < N$. It is worth noting that in the wave functions above $|0\rangle$ refers to the vacuum of $f_x$ operators; i.e., $f_x |0\rangle = 0$. This means that for a given Majorana link between the sites $x$ and $x+1$, we simply multiply the vacuum $|0\rangle$ by the operator $(1 \pm f_x^\dagger)(1 \pm f_{x+1}^\dagger)$. However, these wave functions need to be modified to account for the last link between the end sites $N$ and 1. In the case of an open chain, $g_N^\dagger g_N$ is absent; thus, $|\Psi_+\rangle$ and $|\Psi_-\rangle$ are degenerate. For the periodic boundary condition (PBC), the ground state wave function must be additionally annihilated by $g_N$ operator. Thus, we have

$$\begin{aligned} |\Psi_r\rangle &= g_N |\Psi_+\rangle \\ &= (f_N^\dagger - f_N + f_1^\dagger + f_1)(1 + f_1^\dagger)(1 + f_2^\dagger)\dots(1 + f_N^\dagger)|0\rangle \\ &= |\Psi_+\rangle - |\Psi_-\rangle \\ &= \frac{1}{2^{(N-1)/2}}\prod_{x=1}^{N}(1 + f_x^\dagger)\Big|_{n\in\text{odd}}|0\rangle \end{aligned} \tag{D9}$$

which means only terms with odd number of fermions survive. Thus, the ground state of the PBC $|\Psi_r\rangle$ has odd fermion parity. The anti-periodic boundary condition (APBC) must contain $g_N$ mode, since the Hamiltonian term is $-g_N^\dagger g_N$. Hence, it is

$$\begin{aligned} |\Psi_{ns}\rangle &= g_N^\dagger |\Psi_+\rangle \\ &= (f_N - f_N^\dagger + f_1^\dagger + f_1)(1 + f_1^\dagger)(1 + f_2^\dagger)\dots(1 + f_N^\dagger)|0\rangle \\ &= |\Psi_+\rangle + |\Psi_-\rangle \\ &= \frac{1}{2^{(N-1)/2}}\prod_{x=1}^{N}(1 + f_x^\dagger)\Big|_{n\in\text{even}}|0\rangle \end{aligned} \tag{D10}$$

which means it has even fermion parity.

Let us now compute the partial reflection for the even sector $|\Psi_{ns}\rangle$. The same derivation can be carried out for the odd sector. Consider a chain with $N$ sites in total and $N_\text{part}$ sites in the subsystem (we take $N_\text{part}$ to be even which means reflection with respect to the central link). For simplicity we choose the sites 1 to $N_\text{part}$ to be in the subsystem. The wave function (Eq. (D10)) can be rewritten as

$$|\Psi_{ns}\rangle = \frac{1}{2^{(N-1)/2}}(1 + f_1^\dagger)F_{in}^\dagger(1 + f_{N_\text{part}}^\dagger)F_{out}^\dagger\Big|_{n\in\text{even}}|0\rangle \tag{D11}$$

$$= \frac{1}{2^{(N-1)/2}}[(1 + f_1^\dagger f_{N_\text{part}}^\dagger)F_{in}^{(e)\dagger}F_{out}^{(e)\dagger} + (1 - f_1^\dagger f_{N_\text{part}}^\dagger)F_{in}^{(o)\dagger}F_{out}^{(o)\dagger}]|0\rangle$$

$$+ \frac{1}{2^{(N-1)/2}}[(f_1^\dagger + f_{N_\text{part}}^\dagger)F_{in}^{(e)\dagger}F_{out}^{(o)\dagger} + (f_1^\dagger - f_{N_\text{part}}^\dagger)F_{in}^{(o)\dagger}F_{out}^{(e)\dagger}]|0\rangle \tag{D12}$$

We define the new operators

$$F_{in}^{(o/e)\dagger} = \prod_{j=2}^{N_{\text{part}}-1} (1+f_j^\dagger)\Big|_{n\in \text{odd/even}}$$

$$F_{out}^{(o/e)\dagger} = \prod_{j=N_{\text{part}}+1}^{N} (1+f_j^\dagger)\Big|_{n\in \text{odd/even}} \tag{D13}$$

The partial reflection is defined by

$$\mathcal{R}_{\text{part}} f_j^\dagger \mathcal{R}_{\text{part}}^{-1} = i f_{N_{\text{part}}-(j-1)}^\dagger. \tag{D14}$$

Note that

$$\langle 0| F_{in}^{(o)} \mathcal{R}_{\text{part}} F_{in}^{(o)\dagger} |0\rangle = i\, 2^{N_{\text{part}}-3},$$
$$\langle 0| F_{in}^{(e)} \mathcal{R}_{\text{part}} F_{in}^{(e)\dagger} |0\rangle = 2^{N_{\text{part}}-3}. \tag{D15}$$

So, we can simplify the expression for the partial reflection,

$$Z_N = \langle \Psi_{ns}| \mathcal{R}_{\text{part}} |\Psi_{ns}\rangle = \frac{1}{2^N}\left[2\, 2^{N_{\text{part}}-3} 2^{N-N_{\text{part}}}(1+i) + 2\, 2^{N_{\text{part}}-3} 2^{N-N_{\text{part}}}\, i(1-i)\right]. \tag{D16}$$

Hence, we conclude

$$Z_N = \frac{1+i}{2}. \tag{D17}$$

### Appendix E: Fermionic matrix product states

In this appendix, we review the fermionic matrix product state [15] and show that the partial reflection gives the $\mathbb{Z}_8$ classification as expected for the symmetry class D with reflection. Following Dubail and Read in [15], we write an MPS representation for a fermionic wave function is in terms of Gaussian integrals over Grassmann variables,

$$|\Psi\rangle = \int d[\xi]\; e^{\sum_x \boldsymbol{\xi}^T(x) B \boldsymbol{\xi}(x)}\; e^{\sum_x \boldsymbol{\xi}^T(x+1) A \boldsymbol{\xi}(x)} e^{\sum_x \boldsymbol{\xi}^T(x) \kappa f_x^\dagger} |0\rangle \tag{E1}$$

where $\boldsymbol{\xi}_j^T(x) = (\xi^1(x), \xi^2(x))$ is a two-component Grassmann vector,

$$A = \begin{pmatrix} 1 & \lambda \\ -\lambda & -1 \end{pmatrix}, \tag{E2}$$

$$B = \begin{pmatrix} 0 & -\lambda \\ \lambda & 0 \end{pmatrix}, \tag{E3}$$

and $\kappa^T = (\kappa_1, 0)$. So, the wave function can be written in momentum space as

$$|\Psi\rangle = \int d[\xi]\; e^{\sum_k \boldsymbol{\xi}_{-k}^T S_k \boldsymbol{\xi}_k}\; e^{\sum_k \boldsymbol{\xi}_{-k}^T \kappa f_k^\dagger} |0\rangle \tag{E4}$$

where

$$S_k = \begin{pmatrix} -i\sin k & -\lambda(1-\cos k) \\ \lambda(1-\cos k) & i\sin k \end{pmatrix}, \tag{E5}$$

and the Fourier components are defined by $\xi_k = \frac{1}{\sqrt{N}} \sum_x e^{-ikx} \xi_x$ and $f_k = \frac{1}{\sqrt{N}} \sum_x e^{ikx} f_x$. Integrating $\xi_k$ fields (with anti-periodic boundary condition) leads to

$$|\Psi_{ns}\rangle \propto \exp\left(\sum_{k>0} g_k f_k^\dagger f_{-k}^\dagger\right) |0\rangle \tag{E6}$$

in which

$$g_k = \frac{\kappa_1^2}{2}[S_k^{-1}]_{11} = \frac{\kappa_1^2}{2} \frac{i \sin k}{\sin^2 k + \lambda^2(1-\cos k)^2}. \tag{E7}$$

This is the familiar BCS wave function. However, we should note that the coefficient $g_k$ is not exactly the same as that of the ground state of the BdG Hamiltonian, which is

$$g_k = \frac{i 2\Delta \sin k}{-(2t\cos k + \mu) + \sqrt{(2t\cos k + \mu)^2 + 4\Delta^2 \sin^2 k}}. \tag{E8}$$

Nevertheless, the fermionic MPS defined here is topologically non-trivial and can be used to benchmark the calculations of the partial reflection. In particular, $\lambda = 1$ and $\kappa_1 = 2$ corresponds to the fixed-point limit of the BdG Hamiltonian (with $\mu = 0$ and $t = \Delta$) where

$$g_k = i \cot \frac{k}{2}. \tag{E9}$$

To compute the expectation values, we use the following definition for the complex conjugate wave function

$$\langle \Psi| \equiv (|\Psi\rangle)^\dagger = \int d[\xi]\ \langle 0| e^{\sum_x f_x \kappa^T \xi(x)} e^{\sum_x \xi^T(x) B^\dagger \xi(x)}\ e^{\sum_x \xi^T(x) A^\dagger \xi(x+1)}$$
$$= \int d[\xi]\ \langle 0| e^{\sum_x f_x \kappa^T \xi(x)} e^{\sum_x \xi^T(x) B^\dagger \xi(x)}\ e^{\sum_x \xi^T(x+1)(-A^*) \xi(x)} \tag{E10}$$

Therefore, the wave function overlap ("zero-temperature partition function") is found by

$$\langle \Psi|\Psi\rangle = \int d[\xi] \int d[\eta]\ e^{\sum_x \eta^T(x) B^\dagger \eta(x) + \sum_x \eta^T(x+1)(-A^*)\eta(x)} e^{\sum_x \xi^T(x) B \xi(x) + \sum_x \xi^T(x+1) A \xi(x)} e^{-\sum_x \xi^T(x) \kappa \kappa^T \eta(x)}$$
$$= \int d[\xi] \int d[\eta] \exp\left[(\xi^T, \eta^T) Y \begin{pmatrix} \xi \\ \eta \end{pmatrix}\right]$$
$$= \text{Pf}[Y] \tag{E11}$$

where in the last line, we carry out the integrals over the Grassmann variables. The matrices are given by

$$Y = \begin{pmatrix} M & -K \\ K^T & M^\dagger \end{pmatrix} \tag{E12}$$

in which $M$ defines the hopping between similar sites ($\xi$ or $\eta$),

$$M = \begin{pmatrix} B & A/2 & & & & -A/2 \\ -A/2 & B & A/2 & & & \\ & -A/2 & B & A/2 & & \\ & & \ddots & \ddots & \ddots & \\ & & & -A/2 & B & A/2 \\ A/2 & & & & -A/2 & B \end{pmatrix}, \tag{E13}$$

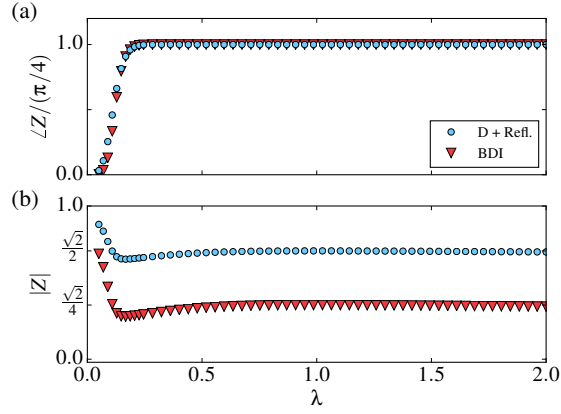

FIG. 2. Partial reflection Eq. (5) and partial transpose Eq. (6) for the fermionic MPS (Eq. (E1)). We set $\kappa_1 = 2$. The total number of sites is $N = 120$ and length of transformed subsystem is $N_{\text{part}} = 60$. For partial transpose, each $I_1$ and $I_2$ has 30 sites.

and $K$ denotes the link between $\xi$ and $\eta$ variables over physical degrees of freedom,

$$K = \frac{1}{2} \begin{pmatrix} \kappa\kappa^T & & & \\ & \kappa\kappa^T & & \\ & & \ddots & \\ & & & \kappa\kappa^T \end{pmatrix}. \tag{E14}$$

Here, the empty matrix entries are zero.

The full reflection operator acts only on the fermion operators $\mathcal{R} f_x \mathcal{R}^{-1} = i f_{-x}$, hence, we have

$$\mathcal{R}|\Psi\rangle = \int d[\xi] \, e^{\sum_x \boldsymbol{\xi}^T(x) B \boldsymbol{\xi}(x)} \, e^{\sum_x \boldsymbol{\xi}^T(x+1) A \boldsymbol{\xi}(x)} e^{-i \sum_x \boldsymbol{\xi}^T(-x) \kappa f_x^\dagger} |0\rangle. \tag{E15}$$

The wave function overlap now becomes

$$\langle \Psi | \mathcal{R}_{\text{part}} | \Psi \rangle = \int d[\xi] \int d[\eta] \, e^{\sum_x \boldsymbol{\eta}^T(x) B^\dagger \boldsymbol{\eta}(x) + \sum_x \boldsymbol{\eta}^T(x+1)(-A^*)\boldsymbol{\eta}(x)} e^{\sum_x \boldsymbol{\xi}^T(x) B \boldsymbol{\xi}(x) + \sum_x \boldsymbol{\xi}^T(x+1) A \boldsymbol{\xi}(x)}$$

$$\times e^{-\sum_{x \in \text{part}} \boldsymbol{\xi}^T(-x) \kappa \kappa^T \boldsymbol{\eta}(x) - \sum_{x \notin \text{part}} \boldsymbol{\xi}^T(x) \kappa \kappa^T \boldsymbol{\eta}(x)}$$

$$= \int d[\xi] \int d[\eta] \exp\left[(\boldsymbol{\xi}^T, \boldsymbol{\eta}^T) \tilde{Y} \begin{pmatrix} \boldsymbol{\xi} \\ \boldsymbol{\eta} \end{pmatrix}\right]$$

$$= \text{Pf}[\tilde{Y}] \tag{E16}$$

where

$$\tilde{Y} = \begin{pmatrix} M & -\tilde{K} \\ \tilde{K}^T & M^\dagger \end{pmatrix} \tag{E17}$$

which has only a different inter-chain linking (as in Fig. 3(a) of the main text):

$$\tilde{K} = \frac{1}{2} \left( \begin{array}{c|c} \begin{matrix} & & i\kappa\kappa^T \\ & \iddots & \\ i\kappa\kappa^T & & \\ i\kappa\kappa^T & & \end{matrix} & \\ \hline & \begin{matrix} \kappa\kappa^T & & \\ & \kappa\kappa^T & \\ & & \ddots \\ & & & \kappa\kappa^T \end{matrix} \end{array} \right). \tag{E18}$$

The upper left block matrix corresponds to the reflected sites and the lower right block corresponds to the unchanged sites. Thus, the normalized partial reflection is given by

$$Z = \frac{\langle\Psi|\mathcal{R}_{\text{part}}|\Psi\rangle}{\langle\Psi|\Psi\rangle} = \frac{\text{Pf}[\tilde{Y}]}{\text{Pf}[Y]}. \tag{E19}$$

The results for various values of $\lambda$ are shown in Fig. 2. For all values $\lambda \neq 0$, the wave function is topological. Looking at $|Z|$ or $\angle Z$, one realizes that there is a smooth transition from $\lambda = 0$ to the finite $\lambda$ due to finite size effects. The plateau of $\angle Z$ at $\pi/4$ matches with the $\mathbb{Z}_8$ classification in the symmetry class D with reflection.

### Appendix F: Partial reflection: More examples

In this appendix, we give three more examples of the partial reflection for a wide range of model parameters. We discuss: 1) Exact ground state wave functions for interacting Majorana chains in class D with reflection, 2) the $s$-wave construction of Majorana chain which is a microscopic model in class D with reflection, and 3) the Su-Shrieffer-Heeger model that is an example of class A with reflection.

#### 1. Exact ground state wave functions for interacting Majorana chains

In this part, we show that the exact ground state wave function of the interacting Majorana chain in class D with reflection (introduced in [16]) can be constructed in terms of fermionic matrix product states discussed in Appendix E. Next, we verify that the $\mathbb{Z}_8$ classification in this case can also be captured by the partial reflection. In addition, we prove this result analytically.

It was shown in [16] that the ground state of the interacting Majorana chain,

$$\hat{H}_{\text{int}} = -\sum_x [tf^\dagger_{x+1}f_x - \Delta f^\dagger_{x+1}f^\dagger_x + \text{H.c.}] - \mu\sum_x f^\dagger_x f_x + U\sum_x (2n_x - 1)(2n_{x+1} - 1), \tag{F1}$$

for parameter values satisfying the condition $\mu = 4\sqrt{U^2 + tU + (t^2 - \Delta^2)/4}$ can be written in a closed form as

$$|\Psi_{r(ns)}\rangle = |\Psi^\alpha_+\rangle \mp |\Psi^\alpha_-\rangle, \tag{F2}$$

where

$$|\Psi^\alpha_\pm\rangle = \frac{1}{(1+\alpha^2)^{N/2}} \prod_{x=1}^{N} (1 \pm \alpha f^\dagger_x)|0\rangle$$
$$= \frac{1}{(1+\alpha^2)^{N/2}} e^{\pm\alpha f^\dagger_1} e^{\pm\alpha f^\dagger_2} \dots e^{\pm\alpha f^\dagger_N}|0\rangle, \tag{F3}$$

corresponding to odd (even) fermion parity sectors associated with (anti-)periodic boundary condition where $\alpha = \sqrt{\cot(\theta/2)}$ and $\theta = \arctan(2\Delta/\mu)$. Using the fact that the two representations of the fixed-point wave function in Eqs. (D10) and (E1) with $\lambda = 1$ and $\kappa_1 = 2$ are identical (see also [17]), it is clear that the interacting wave function given above in Eq. (F2) is equal to the MPS wave function in Eq. (E6) with $\kappa_1 = 2\alpha$. Therefore, we can use our

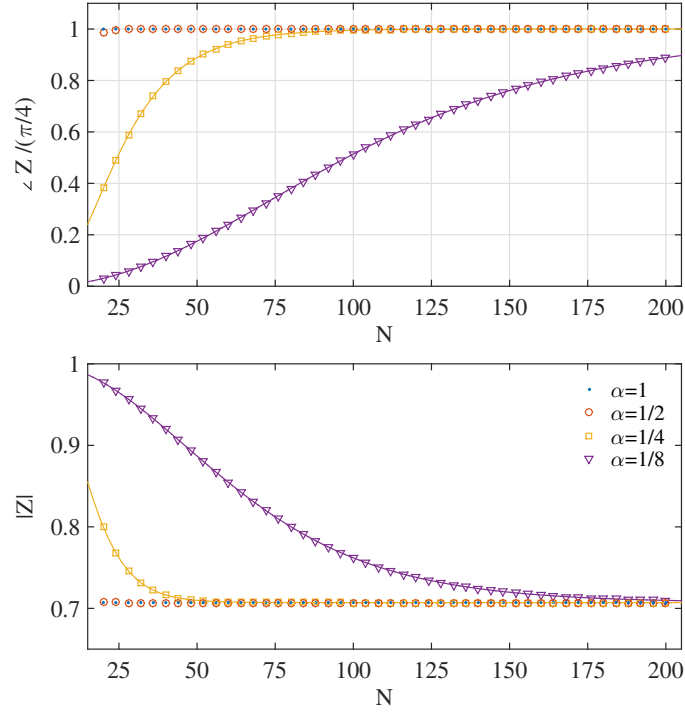

FIG. 3. Partial reflection for the interacting wave function Eq. (F2) as a function of total number of sites $N = 2N_{\text{part}}$ for various values of $\alpha$. $\alpha = 1$ corresponds to the non-interacting fixed-point wave function. Solid lines show analytical expression found in Eq. (F10).

MPS construction of the partial reflection to extract $\mathbb{Z}_8$ phase for generic cases where $\kappa_1 \neq 2$. Figure 3 shows the partial reflection as a function of total number of sites. It is quite remarkable that as we approach the long-chain limit (compared to the correlation length), the amplitudes and complex phases of all various cases converge to the anticipated values $1/\sqrt{2}$ and $\pi/4$, respectively. In the following, we provide an analytical derivation of this result.

We prove our result for the case with anti-periodic boundary condition. A similar derivation can be carried out for the case of periodic boundary condition. Consider a long chain with $N$ sites in total and $N_{\text{part}}$ sites in the subsystem such that $N = 2N_{\text{part}}$ (we take $N_{\text{part}}$ to be even which means reflection with respect to the central link). The wave function is given by

$$|\Psi_{ns}\rangle = \frac{1}{\sqrt{\mathcal{A}_+(N)}} \prod_{x=1}^N (1+\alpha f_x^\dagger)\Big|_{n\in \text{even}} |0\rangle \tag{F4}$$

where the normalization factor is

$$\mathcal{A}_\pm(n) = \frac{1}{2}[(1+\alpha^2)^n \pm (1-\alpha^2)^n]. \tag{F5}$$

For simplicity we choose the sites 1 to $N_{\text{part}}$ to be in the subsystem. The wave function can be rewritten as

$$\begin{aligned}|\Psi_{ns}\rangle =& \frac{1}{\sqrt{\mathcal{A}(N)}}(1+\alpha f_1^\dagger)F_{in}^\dagger(1+\alpha f_{N_{\text{part}}}^\dagger)F_{out}^\dagger\Big|_{n\in\text{even}}|0\rangle \\ =& \frac{1}{\sqrt{\mathcal{A}(N)}}[(1+\alpha^2 f_1^\dagger f_{N_{\text{part}}}^\dagger)F_{in}^{(e)\dagger}F_{out}^{(e)\dagger} + (1-\alpha^2 f_1^\dagger f_{N_{\text{part}}}^\dagger)F_{in}^{(o)\dagger}F_{out}^{(o)\dagger}]|0\rangle \\ &+ \frac{\alpha}{\sqrt{\mathcal{A}(N)}}[(f_1^\dagger + f_{N_{\text{part}}}^\dagger)F_{in}^{(e)\dagger}F_{out}^{(o)\dagger} + (f_1^\dagger - f_{N_{\text{part}}}^\dagger)F_{in}^{(o)\dagger}F_{out}^{(e)\dagger}]|0\rangle. \end{aligned} \tag{F6}$$

We define the new operators

$$F_{in}^{(o/e)\dagger} = \prod_{j=2}^{N_{\text{part}}-1} (1+\alpha f_j^\dagger)\Big|_{n\in\text{odd/even}},$$

$$F_{out}^{(o/e)\dagger} = \prod_{j=N_{\text{part}}+1}^{N} (1+\alpha f_j^\dagger)\Big|_{n\in\text{odd/even}}. \tag{F7}$$

The partial reflection is defined by

$$\mathcal{R}_{\text{part}} f_j^\dagger \mathcal{R}_{\text{part}}^{-1} = i f_{N-(j-1)}^\dagger. \tag{F8}$$

Note that

$$\langle 0| F_{in}^{(o)} \mathcal{R}_{\text{part}} F_{in}^{(o)\dagger} |0\rangle = i\mathcal{A}_-(N_{\text{part}}-2),$$

$$\langle 0| F_{in}^{(o)} \mathcal{R}_{\text{part}} F_{in}^{(o)\dagger} |0\rangle = \mathcal{A}_+(N_{\text{part}}-2). \tag{F9}$$

So, we have

$$Z_N = \langle \Psi_{ns}| \mathcal{R}_{\text{part}} |\Psi_{ns}\rangle = \frac{1}{\mathcal{A}(N)}\left[(1+\alpha^4)\mathcal{A}_+(N_{\text{part}}-2)\mathcal{A}_+(N-N_{\text{part}}) + i(1+\alpha^4)\mathcal{A}_-(N_{\text{part}}-2)\mathcal{A}_-(N-N_{\text{part}})\right]$$

$$+ \frac{2\alpha^2}{\mathcal{A}(N)}\left[i\mathcal{A}_+(N_{\text{part}}-2)\mathcal{A}_-(N-N_{\text{part}}) + \mathcal{A}_-(N_{\text{part}}-2)\mathcal{A}_+(N-N_{\text{part}})\right] \tag{F10}$$

In the thermodynamic limit $N\to\infty$, this becomes

$$\lim_{N\to\infty} Z_N = \frac{(1+\alpha^2)^{N-2}}{2(1+\alpha^2)^N}[(1+\alpha^4)(1+i) + 2\alpha^2(i+1)]$$

$$= \frac{1+i}{2}, \tag{F11}$$

where we use the fact that

$$\lim_{n\to\infty} \mathcal{A}_\pm(n) = \frac{1}{2}(1+\alpha^2)^n. \tag{F12}$$

Figure 3 shows the perfect agreement between analytical results and numerical calculations.

### 2. $s$-wave construction of Majorana chain

As another application of the partial reflection, we study the $s$-wave superconducting nanowire with spin-orbit coupling in the presence of Zeeman field which is proposed as an experimental realization of the Majorana chain [18]. This model is a member of symmetry class D and is equipped with the reflection symmetry. Hence, the interacting classifiction is $\mathbb{Z}_8$. The Hamiltonian is defined by

$$\hat{H} = -t\sum_{x,\sigma}[f_{x+1\sigma}^\dagger f_{x\sigma} + \text{H.c.}] - \mu\sum_{x,\sigma} f_{x\sigma}^\dagger f_{x\sigma}$$

$$+ \Delta\sum_x [f_{x\uparrow}^\dagger f_{x\downarrow}^\dagger + \text{H.c.}] + h_z\sum_{x,\sigma,\sigma'} (s_z)_{\sigma\sigma'} f_{x\sigma}^\dagger f_{x\sigma'}$$

$$+ \lambda\sum_x [f_{x+1\uparrow}^\dagger f_{x\downarrow} - f_{x-1\uparrow}^\dagger f_{x\downarrow} + \text{H.c.}]. \tag{F13}$$

Here, the reflection symmetry acts on fermion operators as $\mathcal{R}f_{x\sigma}\mathcal{R}^{-1} = i(s_z)_{\sigma\sigma'} f_{-x\sigma'}$ where $s_z$ is the third Pauli matrix in the spin basis. We use the partial reflection to characterize each phase and map out the phase diagram as shown in Fig. 4. The phase boundaries are accurately determined. Interestingly, the sign of the topological phase $\pm\frac{\pi}{4}$ at negative (positive) $\mu$ correctly captures the two distinct topological phases associated with $\nu = \pm 1$ corresponding



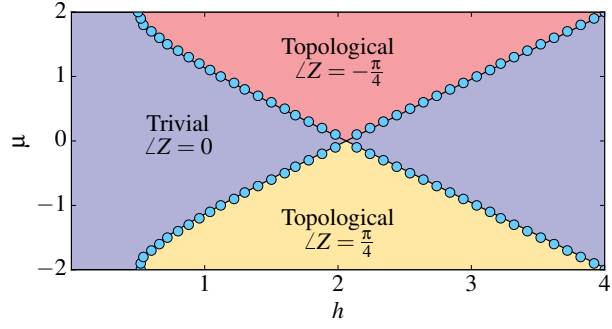

FIG. 4. Phase diagram of $s$-wave superconducting nanowire with spin-orbit coupling (Eq. (F13)) using the partial reflection. The boundary between a topological phase and a trivial phase is identified by a jump in $\angle Z$ from $\pm\pi/4$ to 0. The solid lines show the analytically computed phase boundaries. Here, we set $t=1$ and $\Delta = \lambda = 0.5$. The size of total system and subsystem are $N=200$ and $N_{\text{part}}=100$, respectively.

to two chains with opposite unpaired Majorana modes on each end.

### 3. Class A in the presence of reflection (Class A+R)

Here, we illustrate how the partial reflection can be implemented for class A topological insulators with reflection symmetry. The topological classification is given by the $Pin^c$ bordism in (1+1)d, that is $\Omega_2^{Pin^c} = \mathbb{Z}_4$, where $\mathbb{Z}_4$ is generated by the real projective plane $\mathbb{R}P^2$. We show that by means of the partial reflection we can compute the corresponding topological invariant, both numerically for a wide range of parameters and analytically for the fixed-point wave function.

As a generating model of this symmetry class, we consider the Su-Schrieffer-Heeger (SSH) model in the presence of the reflection symmetry,

$$\hat{H} = -t_2 \sum_x (f_{x+1}^{L\dagger} f_x^R + \text{H.c.}) - t_1 \sum_x (f_x^{L\dagger} f_x^R + \text{H.c.}) \tag{F14}$$

where there are two fermion species living on each site $f_x^L$ and $f_x^R$. This Hamiltonian is symmetric under the reflection which acts on the fermion operators as

$$\mathcal{R} f_x^L \mathcal{R}^{-1} = i f_{-x}^R, \quad \mathcal{R} f_x^R \mathcal{R}^{-1} = i f_{-x}^L. \tag{F15}$$

This model realizes two topologically distinct phases: Topologically non-trivial phase for $t_2 > t_1$, where the open chain has localized fermion modes at the boundaries, and trivial phase for $t_2 < t_1$ which is just an insulator with no boundary mode. Figure 5 shows the complex phase and the amplitude of the partial reflection. Therefore, in the topological phase we have

$$\langle \Psi | \mathcal{R}_{\text{part}} | \Psi \rangle = \frac{e^{i\frac{\pi}{2}}}{2}, \tag{F16}$$

whereas in the trivial phase, we get

$$\langle \Psi | \mathcal{R}_{\text{part}} | \Psi \rangle \leq 1. \tag{F17}$$

This result is consistent with $\mathbb{Z}_4$ classification according to the bordism theory.

Let us now show this result explicitly for the fixed-point wave function when $t_1 = 0$. The ground state is given by

$$|\Psi\rangle = \frac{1}{2^{(N+2)/2}} (f_0^{R\dagger} + f_1^{L\dagger})(f_1^{R\dagger} + f_2^{L\dagger}) \ldots (f_N^{R\dagger} + f_{N+1}^{L\dagger})(f_{N+1}^{R\dagger} + f_0^{L\dagger}) |0\rangle \tag{F18}$$

where we assume ascending order for the fermion indices and take $N$ to be even (hence, the reflection is with respect to the central bond of the subsystem). We define the subsystem as the sites 1 to $N$. We should note that more sites can be included between the sites 0 and $N+1$, but they do not change the final result as there is no entanglement

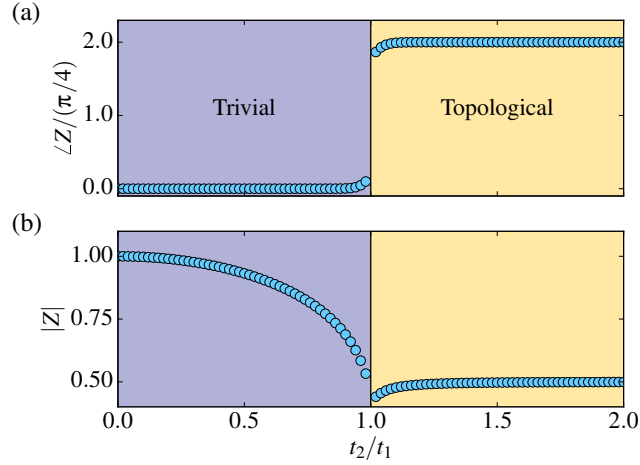

FIG. 5. Partial reflection for the SSH model. Here, $N = 200$ and $N_{\text{part}} = 100$.

between the subsystem and such extra sites. The ground state wave function can be rewritten as

$$|\Psi\rangle = \frac{1}{2^{(N+2)/2}}(f_0^{R\dagger} + f_1^{L\dagger}) F_{in}^\dagger (f_N^{R\dagger} + f_{N+1}^{L\dagger})(f_{N+1}^{R\dagger} + f_0^{L\dagger}) |0\rangle \tag{F19}$$

where

$$F_{in}^\dagger = (f_1^{R\dagger} + f_2^{L\dagger})(f_2^{R\dagger} + f_3^{L\dagger}) \cdots (f_{N-1}^{R\dagger} + f_N^{L\dagger}). \tag{F20}$$

Using the fact that

$$\langle 0| F_{in} \mathcal{R}_{\text{part}} F_{in}^\dagger |0\rangle = 2^{N-1} i^{N-1} (-1)^{(N-1)(N-2)/2}, \tag{F21}$$

we arrive at

$$\begin{aligned}
Z_N &= \langle \Psi | \mathcal{R}_{\text{part}} | \Psi \rangle \\
&= \frac{2^{N-1} i^{N-1}}{2^{N+1}} (-1)^{(N-1)(N-2)/2} \, \langle 0| (f_N^R + f_{N+1}^L)(f_0^R + f_1^L)(f_0^{R\dagger} + i f_N^{R\dagger})(f_{N+1}^{L\dagger} + i f_1^{L\dagger}) |0\rangle \\
&= \frac{i^{(N-1)^2}}{2} = \frac{i}{2}.
\end{aligned} \tag{F22}$$

The last equality holds since $N$ is even. Therefore, we recover the $\mathbb{Z}_4$ classification. We should note that here the amplitude is $\frac{1}{2}$ in contrast to that of the Majorana chain which is $\frac{1}{\sqrt{2}}$. In other words, the amplitude is related to the quantum dimension of the boundary modes in the form of $\prod_j d_j^{-1}$ where $d_i$ is the quantum dimension of the $j$th boundary mode.

### Appendix G: Partial transpose: More details

In this appendix, we present our definition of the partial transpose for a fermionic density matrix to simulate and compute a time-reversal cross-cap in terms of wave function overlaps. The strategy of constructing the fermionic partial transpose developed in this paper is to reproduce the complex $\mathbb{Z}_8$ phase in the partition function of spin topological quantum field theory on real projective plane $\mathbb{R}P^2$ [13]. The resulting definition of the fermionic partial transpose differs from the one that was originally introduced in Ref. [19].

After reviewing the partial transpose of a bosonic density matrix in Sec. G 1, we define the fermionic partial transpose in the coherent state basis inspired by our construction of time-reversal cross-cap in the path integral formalism, in Sec. G 2. In Sec. G 3, we review the general form of the fermionic partial transpose in terms of Majorana operators

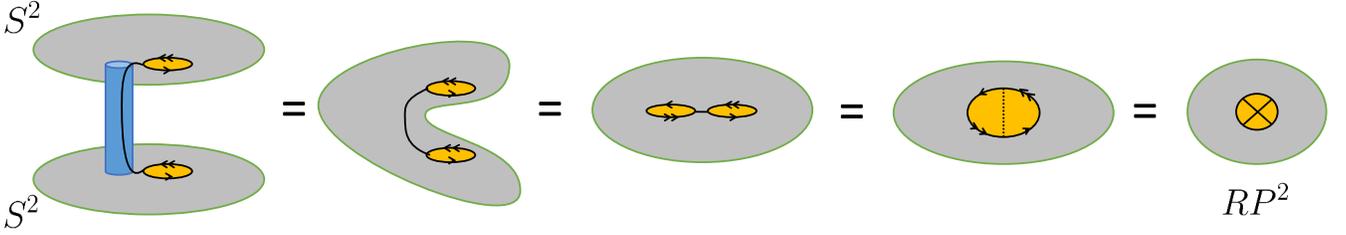

FIG. 6. Equivalence between $Z_\mathcal{T} = \text{tr}(\rho_I U_{I_1} \rho_I^{T_1} U_{I_1}^\dagger)$ and the partition function over the real projective plane $RP^2$. Black lines with the same arrows are identified.

and show that both definition using the coherent state basis and Majorana operators are equivalent as long as we fix $\theta = \pi/2$ for the complex phase in the definition of a transposed Majorana operator. After we establish our definition of the partial transpose, we examine it for the fermionic MPS (Sec. G 4) and the BdG ground state wave functions (Sec. G 5). In Sec. G 6, we analytically derive the partial transpose for the fixed-point wave function of the Majorana chain and also check that the complex phase of the partial transpose is additive as we introduce more fermion flavors In all cases, we arrive at the $\mathbb{Z}_8$ classification corresponding to the symmetry class BDI.

### 1. Partial transpose for bosonic density matrix

Before we get into details of calculations, let us briefly review the definition of the partial transpose of a bosonic density matrix. For a bosonic system (e.g., spin chains), a reduced density matrix $\rho_I = \text{tr}_{S \setminus I}(\rho)$ of a subsystem $I = I_1 \cup I_2$ consisting of two smaller subsystems $I_1$ and $I_2$ can be written as

$$\rho_I = \sum_{ijkl} \langle e_i^1, e_j^2 | \rho_I | e_k^1, e_l^2 \rangle \, |e_i^1, e_j^2\rangle \langle e_k^1, e_l^2|, \tag{G1}$$

where $|e_j^1\rangle$ and $|e_k^2\rangle$ denote orthonormal bases on the Hilbert spaces $\mathcal{H}_1$ and $\mathcal{H}_2$ corresponding to the $I_1$ and $I_2$ regions. The partial transpose of a density matrix for the subsystem $I_1$ is defined by exchanging the matrix elements in the subsystem $I_1$ as in

$$\rho_I^{T_1} = \sum_{ijkl} \langle e_k^1, e_j^2 | \rho_I | e_i^1, e_l^2 \rangle \, |e_i^1, e_j^2\rangle \langle e_k^1, e_l^2|, \tag{G2}$$

which is equivalent to the following transformation on the operator basis

$$\left( |e_i^1, e_j^2\rangle \langle e_k^1, e_l^2| \right)^{T_1} = |e_k^1, e_j^2\rangle \langle e_i^1, e_l^2|. \tag{G3}$$

### 2. Partial transpose in fermionic coherent state representation

As we have seen in the main text and Appendix B, a time-reversal cross-cap can be implemented in the spacetime path integral by rearranging the Grassmann variables and the partition function of a Majorana chain with time-reversal symmetry in the presence of a cross-cap yields the $\mathbb{Z}_8$ complex phase associated with the $\mathbb{Z}_8$ classification of class BDI (see Eqs. (6) and (7) of main text). This motivates us to adapt this procedure for the ground state wave function and define the quantity $Z_\mathcal{T} = \text{tr}(\rho_I U_{I_1} \rho_I^{T_1} U_{I_1}^\dagger)$ where $\rho_I^{T_1}$ involves the fermionic partial transpose. It is worth noting that the quantity $Z_\mathcal{T}$ for adjacent intervals is topologically equivalent to the partition function on $\mathbb{R}P^2$ (as depicted in Fig. 6) [9]. However, if we choose disjoint intervals, such that $I_1 \cap I_2 = \emptyset$, it becomes equivalent to the partition function over the Klein bottle [20].

In order to numerically compute $Z_\mathcal{T}$ for a given ground state wave function, we use the coherent state representation of fermions in terms of Grassmann variables. As we show in the following, the partial transpose can be implemented as a transformation in the space of Grassmann variables. Later, we illustrate this idea for two examples: Fermionic MPS (E1) and BdG ground state (G33). As we will see, the transformation of Grassmann variables here is very similar to that of partition function discussed in Appendix B for the partial time inversion in spacetime.

Let us begin by noting that any density matrix can be expanded in a coherent state basis as

$$\rho = \int d[\bar{\xi}]d[\xi]d[\bar{\chi}]d[\chi] e^{-\sum_j (\bar{\xi}_j \xi_j + \bar{\chi}_j \chi_j)} \langle\{\bar{\xi}_j\}|\rho|\{\chi_j\}\rangle |\{\xi_j\}\rangle \langle\{\bar{\chi}_j\}|, \tag{G4}$$

where $\xi_j, \bar{\xi}_j, \chi_j$ and $\bar{\chi}_j$ are independent Grassmann variables which anticommute with fermion creation/annihilation operators, and $|\{\xi_j\}\rangle := e^{-\sum_j \xi_j f_j^\dagger}|0\rangle, \langle\{\bar{\chi}_j\}| := \langle 0| e^{-\sum_j f_j \bar{\chi}_j}$ are the fermionic coherent states. Note that $|\{\xi_j\}\rangle, \langle\{\bar{\chi}_j\}|$ are Grassmann even and commute with a single Grassmann variable.

Recall that for class BDI, the time-reversal symmetry (TRS) operation is defined by

$$\mathcal{T} f_j^\dagger \mathcal{T}^{-1} = f_j^\dagger, \qquad\qquad \mathcal{T} i \mathcal{T}^{-1} = -i. \tag{G5}$$

The TRS transformation of Grassmann variables in the path integral formalism, defined in the main text, suggests that the fermionic partial transpose in the coherent basis is given by

$$U_{I_1} \left( |\{\xi_j\}_{j \in I_1}, \{\xi_j\}_{j \in I_2}\rangle \langle\{\bar{\chi}_j\}_{j \in I_1}, \{\bar{\chi}_j\}_{j \in I_2}| \right)^{T_1} U_{I_1}^\dagger := |\{-i\bar{\chi}_j\}_{j \in I_1}, \{\xi_j\}_{j \in I_2}\rangle \langle\{-i\xi_j\}_{j \in I_1}, \{\bar{\chi}_j\}_{j \in I_2}| \tag{G6}$$

in a similar manner to (G3). Let us focus only on the combined operation (G6) at the present stage. We will discuss the operator formalism and in particular the unitary operator $U_{I_1}$ in Sec. G 3. There, we explain the partially transposed operator $O^{T_1}$ where $O$ is a generic Grassmann even operator acting on the Fock space.

It is easy to see that the definition (G6) effectively acts as a "partial transposition" similar to (G3) in the occupation number basis. The occupation number basis is defined by

$$|\{n_j\}_{j \in I_1}, \{n_j\}_{j \in I_2}\rangle := (f_1^\dagger)^{n_1} \cdots (f_{M_1}^\dagger)^{n_{M_1}} (f_{M_1+1}^\dagger)^{n_{M_1+1}} \cdots (f_N^\dagger)^{n_N} |0\rangle \tag{G7}$$

for the subsystem $I = I_1 \cup I_2 = \{1, \ldots, M_1\} \cup \{M_1 + 1, \ldots, M\}$. For example, let us consider a two complex fermion system. The operator basis in the coherent state is

$$\begin{aligned}
|\xi_1, \xi_2\rangle \langle \bar{\chi}_1, \bar{\chi}_2| &= |00\rangle \langle 00| + \bar{\chi}_1 |00\rangle \langle 10| + \bar{\chi}_2 |00\rangle \langle 01| + \bar{\chi}_1 \bar{\chi}_2 |00\rangle \langle 11| \\
&\quad - \xi_1 |10\rangle \langle 00| + \xi_1 \bar{\chi}_1 |10\rangle \langle 10| + \xi_1 \bar{\chi}_2 |10\rangle \langle 01| - \xi_1 \bar{\chi}_1 \bar{\chi}_2 |10\rangle \langle 11| \\
&\quad - \xi_2 |01\rangle \langle 00| - \bar{\chi}_1 \xi_2 |01\rangle \langle 10| + \xi_2 \bar{\chi}_2 |01\rangle \langle 01| + \bar{\chi}_1 \xi_2 \bar{\chi}_2 |01\rangle \langle 11| \\
&\quad - \xi_1 \xi_2 |11\rangle \langle 00| + \xi_1 \bar{\chi}_1 \xi_2 |11\rangle \langle 10| - \xi_1 \xi_2 \bar{\chi}_2 |11\rangle \langle 01| + \xi_1 \bar{\chi}_1 \xi_2 \bar{\chi}_2 |11\rangle \langle 11|.
\end{aligned} \tag{G8}$$

Thus, the transformation (G6) is given by

$$\begin{aligned}
U_{I_1} \left( |\xi_1, \xi_2\rangle \langle \bar{\chi}_1, \bar{\chi}_2| \right)^{T_1} U_{I_1}^\dagger &= |-i\bar{\chi}_1, \xi_2\rangle \langle -i\xi_1, \bar{\chi}_2| \\
&= |00\rangle \langle 00| + i\bar{\chi}_1 |10\rangle \langle 00| + \bar{\chi}_2 |00\rangle \langle 01| - i\bar{\chi}_1 \bar{\chi}_2 |10\rangle \langle 01| \\
&\quad - i\xi_1 |00\rangle \langle 10| + \xi_1 \bar{\chi}_1 |10\rangle \langle 10| - i\xi_1 \bar{\chi}_2 |00\rangle \langle 11| - \xi_1 \bar{\chi}_1 \bar{\chi}_2 |10\rangle \langle 11| \\
&\quad - \xi_2 |01\rangle \langle 00| + i\bar{\chi}_1 \xi_2 |11\rangle \langle 00| + \xi_2 \bar{\chi}_2 |01\rangle \langle 01| + i\bar{\chi}_1 \xi_2 \bar{\chi}_2 |11\rangle \langle 01| \\
&\quad + i\xi_1 \xi_2 |01\rangle \langle 10| + \xi_1 \bar{\chi}_1 \xi_2 |11\rangle \langle 10| - i\xi_1 \xi_2 \bar{\chi}_2 |01\rangle \langle 11| + \xi_1 \bar{\chi}_1 \xi_2 \bar{\chi}_2 |11\rangle \langle 11|.
\end{aligned} \tag{G9}$$

Comparing coefficients of these terms, the transformation (G6) can be read in the form of

$$|nm\rangle \langle nm'| \mapsto |nm\rangle \langle nm'|, \quad (n, m, m' \in \{0, 1\}). \tag{G10}$$

for the diagonal terms in the basis of the first fermion and in the following forms for the other terms,

$$\begin{array}{llll}
|00\rangle \langle 10| \mapsto i |10\rangle \langle 00|, & |00\rangle \langle 11| \mapsto -i |10\rangle \langle 01|, & |10\rangle \langle 00| \mapsto i |00\rangle \langle 10|, & |10\rangle \langle 01| \mapsto -i |00\rangle \langle 11|, \\
|01\rangle \langle 10| \mapsto -i |11\rangle \langle 00|, & |01\rangle \langle 11| \mapsto i |11\rangle \langle 01|, & |11\rangle \langle 00| \mapsto -i |01\rangle \langle 10|, & |11\rangle \langle 01| \mapsto i |01\rangle \langle 11|.
\end{array} \tag{G11}$$

To understand how this procedure generalizes to bigger systems, it is easier to work with operators. It is important

to note that the operator basis in the fermionic coherent state basis factorizes such that,

$$|\{\xi_j\}\rangle\langle\{\bar{\chi}_j\}| = \prod_j \left[e^{-\xi_j f_j^\dagger}(1 - f_j^\dagger f_j)e^{-f_j\bar{\chi}_j}\right]$$

$$= \prod_j \left[1 - f_j^\dagger f_j - \xi_j f_j^\dagger + \bar{\chi}_j f_j + \xi_j \bar{\chi}_j f_j^\dagger f_j\right], \tag{G12}$$

and hence, the transformation (G6) reads

$$U_{I_1}\left(|\{\xi_j\}_{j\in I_1},\{\xi_j\}_{j\in I_2}\rangle\langle\{\bar{\chi}_j\}_{j\in I_1},\{\chi_j\}_{j\in I_2}|\right)^{T_1} U_{I_1}^\dagger$$
$$= \prod_{j\in I_1} \left[1 - f_j^\dagger f_j - i\xi_j f_j + i\bar{\chi}_j f_j^\dagger + \xi_j \bar{\chi}_j f_j^\dagger f_j\right] \cdot \prod_{j\in I_2} \left[1 - f_j^\dagger f_j - \xi_j f_j^\dagger + \bar{\chi}_j f_j + \xi_j \bar{\chi}_j f_j^\dagger f_j\right]. \tag{G13}$$

The above expression is equivalent to the simultaneous transformations

$$1 - f_j^\dagger f_j \mapsto 1 - f_j^\dagger f_j, \qquad f_j^\dagger f_j \mapsto f_j^\dagger f_j, \qquad f_j^\dagger \mapsto if_j, \qquad f_j \mapsto if_j^\dagger, \qquad \text{for } j \in I_1, \tag{G14}$$

which have the following form in the occupation number basis

$$|\{n_j\}_{j\in I_1},\{n_j\}_{j\in I_2}\rangle\langle\{\bar{n}_j\}_{j\in I_1},\{\bar{n}_j\}_{j\in I_2}| \mapsto e^{i\phi(\{n_j\},\{\bar{n}_j\})}|\{\bar{n}_j\}_{j\in I_1},\{n_j\}_{j\in I_2}\rangle\langle\{n_j\}_{j\in I_1},\{\bar{n}_j\}_{j\in I_2}|, \quad n_j,\bar{n}_j \in \{0,1\}, \tag{G15}$$

with the $U(1)$ phase $e^{i\phi(\{n_j\},\{\bar{n}_j\})}$ determined by (G13). We should note that (G15) is the same as the bosonic partial transpose (G15) up to the presence of the $U(1)$ phase factor.

### 3. Partial transpose on fermionic operators

Here, we define the partial transpose on fermionic operators in the Fock space. Our goal is to determine how Majorana operators transform under the transformation rule defined by (G6) in the coherent state basis. This gives us an idea about what this transformation does compared to the known results in the literature for the partial transpose, e.g. Ref. [19]. Before we proceed, note that the transformation in (G6) consists of two steps: Taking the partial transpose $\rho_I^{T_1}$ and applying the unitary transformation $U_{I_1}\rho_I^{T_1}U_{I_1}^\dagger$. We will show what each step does to Majorana fermions.

Consider a system of $N$ fermionic states. We introduce Majorana fermions

$$c_{2j-1} = f_j + f_j^\dagger, \quad c_{2j} = i(f_j - f_j^\dagger). \tag{G16}$$

Let

$$\rho_I = \sum_{\kappa,\tau,|\kappa|+|\tau|=\text{even}} w_{\kappa,\tau} c_{m_1}^{\kappa_1} \cdots c_{m_{2k}}^{\kappa_{2k}} c_{n_1}^{\tau_1} \cdots c_{n_{2l}}^{\tau_{2l}} \tag{G17}$$

be the density matrix in the Majorana representation for the subsystem $I = I_1 \cup I_2 = \{m_1,\ldots,m_{2k}\} \cup \{n_1,\ldots,n_{2l}\}$. We use the notation $c_x^0 = \mathbb{I}$ and $c_x^1 = c_x$, so that $\kappa_i, \tau_j \in \{0,1\}$. We also introduce the vectors (bit strings) $\kappa = (\kappa_1,\ldots\kappa_{2k})$, $\tau = (\tau_1,\ldots,\tau_{2l})$, and their norms $|\kappa| = \sum_{j=1}^{2k}\kappa_j, |\tau| = \sum_{j=1}^{2l}\tau_j$. We assume the density matrix $\rho_I$ commutes with the fermion parity operator $(-1)^{F_I} = \prod_{j=1}^k ic_{n_{2j-1}}c_{n_{2j}} \prod_{j=1}^l ic_{m_{2j-1}}c_{m_{2j}}$ on the subsystem $I$, which implies that the sum is over all possible bit strings provided that the constraint $|\kappa| + |\tau| = $ even is satisfied. Using this expression of the density matrix, we introduce the partial transpose for the subsystem $I_1$ in a similar fashion to Ref. [19],

$$\rho_I^{T_1} := \sum_{\kappa,\tau,|\kappa|+|\tau|=\text{even}} w_{\kappa,\tau} \mathcal{R}(c_{m_1}^{\kappa_1}\cdots c_{m_{2k}}^{\kappa_{2k}}) c_{n_1}^{\tau_1}\cdots c_{n_{2l}}^{\tau_{2l}}. \tag{G18}$$

In the following, we show that the definition (G6) leads to the condition $\mathcal{R}(M_1 M_2) = \mathcal{R}(M_1)\mathcal{R}(M_2)$ for two operators $M_1$ and $M_2$ acting on the subsystem $I_1$, which contrasts with Ref. [19] where $\mathcal{R}$ is introduced to satisfy $\mathcal{R}(M_1 M_2) = \mathcal{R}(M_2)\mathcal{R}(M_1)$. Moreover, we see that Eq. (G6) dictates the choice $e^{i\theta} = \pm i$ for the $U(1)$ phase freedom in the partial

transpose $\mathscr{R}(c) = e^{i\theta} c$ for a single Majorana fermion.

First, we need to introduce a TRS-dependent unitary operator $U_{I_1}$ on the subsystem $I_1$. For the symmetry class BDI, TRS is defined in (G5) and the TRS operator is given by $\mathcal{T} = (\prod_j c_{2j-1})\mathcal{K}$ where $\mathcal{K}$ is the complex conjugation. Therefore, we define $U_{I_1}$ as a unitary part acting on the subsystem $I_1$,

$$U_{I_1} := \prod_{j \in I_1} c_{2j-1}. \tag{G19}$$

Second, in order to determine the action of $\mathscr{R}$ on Majorana operators, we compute both sides of (G6) separately and finally identify them. From the left-hand side of (G6), we get

$$U_{I_1} \big( |\{\xi_j\}\rangle \langle \{\bar{\chi}_j\}| \big)^{T_1} U_{I_1}^\dagger$$
$$= U_{I_1} \Big[ \prod_{j \in I} \Big( \frac{1 + \xi_j \bar{\chi}_j}{2} - \frac{\xi_j - \bar{\chi}_j}{2} c_{2j-1} - i \frac{\xi_j + \bar{\chi}_j}{2} c_{2j} + \frac{i(1 - \xi_j \bar{\chi}_j)}{2} c_{2j-1} c_{2j} \Big) \Big]^{T_1} U_{I_1}^\dagger$$
$$= \prod_{j \in I_1} c_{2j-1} \mathscr{R} \Big[ \prod_{n \in I_1} \Big( \frac{1 + \xi_n \bar{\chi}_n}{2} - \frac{\xi_n - \bar{\chi}_n}{2} c_{2n-1} - i \frac{\xi_n + \bar{\chi}_n}{2} c_{2n} + \frac{i(1 - \xi_n \bar{\chi}_n)}{2} c_{2n-1} c_{2n} \Big) \Big] \overset{\leftarrow}{\prod_{j \in I_1}} c_{2j-1}$$
$$\cdot \prod_{n \in I_2} \Big( \frac{1 + \xi_n \bar{\chi}_n}{2} - \frac{\xi_n - \bar{\chi}_n}{2} c_{2n-1} - i \frac{\xi_n + \bar{\chi}_n}{2} c_{2n} + \frac{i(1 - \xi_n \bar{\chi}_n)}{2} c_{2n-1} c_{2n} \Big). \tag{G20}$$

where $U_{I_1}^\dagger = \overset{\leftarrow}{\prod_{j \in I_1}} c_{2j-1}$ which means that Majorana operators are arranged in the reversed order compared to that of $U_{I_1}$. The right-hand side of (G6) also yields

$$|\{-i\bar{\chi}_j\}_{j \in I_1}, \{\xi_j\}_{j \in I_2}\rangle \langle \{-i\xi_j\}_{j \in I_1}, \{\bar{\chi}_j\}_{j \in I_2}|$$
$$= \prod_{j \in I_1} \Big( \frac{1 + \xi_j \bar{\chi}_j}{2} - \frac{i\xi_j - i\bar{\chi}_j}{2} c_{2j-1} - i \frac{-i\xi_j - i\bar{\chi}_j}{2} c_{2j} + \frac{i(1 - \xi_j \bar{\chi}_j)}{2} c_{2j-1} c_{2j} \Big)$$
$$\cdot \prod_{j \in I_2} \Big( \frac{1 + \xi_j \bar{\chi}_j}{2} - \frac{\xi_j - \bar{\chi}_j}{2} c_{2j-1} - i \frac{\xi_j + \bar{\chi}_j}{2} c_{2j} + \frac{i(1 - \xi_j \bar{\chi}_j)}{2} c_{2j-1} c_{2j} \Big). \tag{G21}$$

Now, we compare the left-hand side (G20) and the right-hand side (G21) of Eq. (G6) by identifying the operators multiplying the same Grassmann variables. This identification in turn fixes the partial transpose $\mathscr{R}(c) = -ic$ and enforces the condition $\mathscr{R}(M_1 M_2) = \mathscr{R}(M_1)\mathscr{R}(M_2)$ (i.e., no reversed ordering). It is important to note that if we had reversed ordering, the right-hand side (G21) must have had terms like $\xi_{n-j} c_{2j-1}$ (i.e., we should have had reversed ordering of Grassmann variables in (G21) besides exchanging them from bra to ket states).

So far, we have shown that the cross-cap as defined by the path-integral can be recast as a change of Grassmann variables in the density matrix written in the coherent state basis. We have also found the equivalent transformation rules governed by $U_{I_1} \rho_I^T U_{I_1}^\dagger$ for the Majorana operators. In summary, the key equation of this section is

$$U_{I_1} \rho_I^{T_1} U_{I_1}^\dagger = \int d[\bar{\xi}] d[\xi] d[\bar{\chi}] d[\chi] e^{-\sum_j (\bar{\xi}_j \xi_j + \bar{\chi}_j \chi_j)} |\{-i\bar{\chi}_j\}_{j \in I_1}, \{\xi_j\}_{j \in I_2}\rangle \rho_I(\{\bar{\xi}_j\};\{\chi_j\}) \langle \{-i\xi_j\}_{j \in I_1}, \{\bar{\chi}_j\}_{j \in I_2}|. \tag{G22}$$

where $\rho_I(\{\bar{\xi}_j\};\{\chi_j\}) = \langle \{\bar{\xi}_j\}| \rho_I |\{\chi_j\}\rangle$. The above expression is then used to numerically evaluate the topological invariant $Z_\mathcal{T} = \text{tr}(\rho_I U_{I_1} \rho_I^{T_1} U_{I_1}^\dagger)$ in the following sections.

### 4. Fermionic matrix product state

Here, we show how we compute the partial transpose for the fermionic MPS introduced in Eq. (E1). Consider a chain with $N$ sites where $I_1$ corresponds to the sites $1, \ldots, M_1$ and $I_2$ corresponds to the sites $M_1 + 1, \ldots, M_1 + M_2 = M$.

First, we need to compute the reduced density matrix, that is found to be

$$\begin{aligned} \rho_I &= \mathrm{tr}_{S \setminus I} (|\Psi\rangle \langle \Psi|) \\ &= \int d[\xi] d[\bar{\xi}] \ e^{\sum_{xx'} \bar{\xi}^T(x) M^\dagger_{xx'} \bar{\xi}(x')} e^{\sum_{xx'} \xi^T(x) M_{xx'} \xi(x')} \\ &\quad \times e^{-\sum_{x \notin I} \bar{\xi}^T(x) \kappa \kappa^T \xi(x)} e^{\sum_{x \in I} \xi^T(x) \kappa f^\dagger_x} |0\rangle \langle 0| e^{\sum_{x \in I} f_x \kappa^T \bar{\xi}(x)}. \end{aligned} \quad (G23)$$

where the matrix $M$ is defined in Eq. (E13). Its partial transpose then becomes

$$\begin{aligned} U_{I_1} \rho_I^{T_1} U_{I_1}^\dagger &= \int d[\xi] d[\bar{\xi}] \ e^{\sum_{xx'} \bar{\xi}^T(x) M^\dagger_{xx'} \bar{\xi}(x')} e^{\sum_{xx'} \xi^T(x) M_{xx'} \xi(x')} \\ &\quad \times e^{-\sum_{x \notin I} \bar{\xi}^T(x) \kappa \kappa^T \xi(x)} e^{-\sum_{x \in I_1} i\bar{\xi}^T(x) \kappa f^\dagger_x + \sum_{x \in I_2} \xi^T(x) \kappa f^\dagger_x} |0\rangle \langle 0| e^{-\sum_{x \in I_1} i f_x \kappa^T \xi(x) + \sum_{x \in I_2} f_x \kappa^T \bar{\xi}(x)} \end{aligned} \quad (G24)$$

Therefore, the overlap can be computed as follows

$$\begin{aligned} Z_{\mathcal{T}} &= \mathrm{tr}(\rho_I U_{I_1} \rho_I^{T_1} U_{I_1}^\dagger) \\ &= \int d[\xi] d[\bar{\xi}] \int d[\eta] d[\bar{\eta}] \ e^{\sum_{xx'} \bar{\xi}^T(x) M^\dagger_{xx'} \bar{\xi}(x')} e^{\sum_{xx'} \xi^T(x) M_{xx'} \xi(x')} e^{\sum_{xx'} \bar{\eta}^T(x) M^\dagger_{xx'} \bar{\eta}(x')} e^{\sum_{xx'} \eta^T(x) M_{xx'} \eta(x')} \\ &\quad \times e^{-\sum_{x \notin I} \bar{\xi}^T(x) \kappa \kappa^T \xi(x)} e^{-\sum_{x \notin I} \bar{\eta}^T(x) \kappa \kappa^T \eta(x)} \\ &\quad \times e^{-\sum_{x \in I_1} i\bar{\xi}^T(x) \kappa \kappa^T \bar{\eta}(x) + \sum_{x \in I_2} \xi^T(x) \kappa \kappa^T \bar{\eta}(x)} e^{-\sum_{x \in I_1} i\eta(x) \kappa \kappa^T \xi(x) + \sum_{x \in I_2} \eta(x) \kappa \kappa^T \bar{\xi}(x)} \\ &= \int d[\xi] d[\bar{\xi}] \int d[\eta] d[\bar{\eta}] \exp\left[ (\xi^T, \bar{\xi}^T, \eta^T, \bar{\eta}^T) \ \tilde{W} \ (\xi^T, \bar{\xi}^T, \eta^T, \bar{\eta}^T)^T \right] \quad (G25) \\ &= \mathrm{Pf}[\tilde{W}]. \quad (G26) \end{aligned}$$

The matrix $\tilde{W}$ is given by

$$\tilde{W} = \begin{pmatrix} \tilde{Y} & -V \\ V^T & \tilde{Y} \end{pmatrix}, \quad (G27)$$

where $\tilde{Y}$ and V are a $2N \times 2N$ matrix defined by

$$\tilde{Y} = \begin{pmatrix} M & -K_1 \\ K_1^T & M^\dagger \end{pmatrix}, \quad (G28)$$

and

$$V = \begin{pmatrix} K_2 & K_3 \\ K_3 & K_2 \end{pmatrix}. \quad (G29)$$

In other words, $\tilde{Y}$ describes the intra-ladder (blue) links and V describes inter-ladder (green and purple) links in Fig. 3(b) of the main text. The block matrices inside $\tilde{Y}$ and V also take a block diagonal form in $\{I_1, I_2, S \setminus I\}$ basis

as in

$$K_1 = \frac{1}{2} \begin{pmatrix} 0 & & & & & & & & & \\ & 0 & & & & & & & & \\ & & \ddots & & & & & & & \\ & & & 0 & & & & & & \\ \hline & & & & 0 & & & & & \\ & & & & & 0 & & & & \\ & & & & & & \ddots & & & \\ & & & & & & & 0 & & \\ \hline & & & & & & & & \kappa\kappa^T & & \\ & & & & & & & & & \kappa\kappa^T & \\ & & & & & & & & & & \ddots & \\ & & & & & & & & & & & \kappa\kappa^T \end{pmatrix}, \tag{G30}$$

$$K_2 = \frac{1}{2} \begin{pmatrix} -i\kappa\kappa^T & & & & & & & & \\ & -i\kappa\kappa^T & & & & & & & \\ & & \ddots & & & & & & \\ & & & -i\kappa\kappa^T & & & & & \\ \hline & & & & 0 & & & & \\ & & & & & 0 & & & \\ & & & & & & \ddots & & \\ & & & & & & & 0 & \\ \hline & & & & & & & & 0 & \\ & & & & & & & & & 0 \\ & & & & & & & & & & \ddots \\ & & & & & & & & & & & 0 \end{pmatrix}, \tag{G31}$$

and

$$K_3 = \frac{1}{2} \begin{pmatrix} 0 & & & & & & & & \\ & 0 & & & & & & & \\ & & \ddots & & & & & & \\ & & & 0 & & & & & \\ \hline & & & & \kappa\kappa^T & & & & \\ & & & & & \kappa\kappa^T & & & \\ & & & & & & \ddots & & \\ & & & & & & & \kappa\kappa^T & \\ \hline & & & & & & & & 0 \\ & & & & & & & & & 0 \\ & & & & & & & & & & \ddots \\ & & & & & & & & & & & 0 \end{pmatrix}. \tag{G32}$$

$K_1$, $K_2$, and $K_3$ correspond to the blue, green and purple links in Fig. 3 of main text, respectively. Note that only links in $K_2$ are multiplied by a factor of $-i$. Figure 2 shows the partial reflection and partial transpose of the fermionic MPS over a range of $\lambda$ values. Clearly, they match the results in the main text. Note that for any $\lambda \neq 0$ the above wave function is topological. In particular, $\lambda = 1$ corresponds to the fixed-point wave function and $\lambda = 0$ corresponds to the critical phase. Hence, as $\lambda \to 0$, the correlation length diverges. As a result, there is a smooth transition in our finite size calculations.

### 5. Grassmann representation of BdG ground state

In Appendix C, we show that the ground state of the BdG Hamiltonian can be written in the form of Eq. (C8) in terms of the matrix $\phi_{ij}$. Inspired by the results of fermionic MPS in the previous appendix, we can reverse engineer the process and introduce the wave function

$$|\Psi\rangle = \int d[\xi] \; e^{\sum_{ij} \xi_i [\phi^{-1}]_{ij} \xi_j} \; e^{\sum_j \xi_j f_j^\dagger} |0\rangle. \tag{G33}$$

It is straightforward to see that the above wave function is identical to the BdG wave function (Eq. (C8)) as one carries out the Gaussian integral over $\xi_j$ variables. Alternatively, we can start by representing the wave function in the coherent state basis

$$\begin{aligned}
|\Psi\rangle &= \int d[\xi] d[\bar{\xi}] e^{-\sum_j \bar{\xi}_j \xi_j} \langle \xi | \Psi \rangle |\xi\rangle \\
&= \int d[\xi] d[\bar{\xi}] e^{-\sum_j \bar{\xi}_j \xi_j} \langle \bar{\xi} | e^{\sum_{ij} f_i^\dagger \phi_{ij} f_j^\dagger} |0\rangle |\xi\rangle \\
&= \int d[\xi] d[\bar{\xi}] e^{-\sum_j \bar{\xi}_j \xi_j} e^{\sum_{ij} \bar{\xi}_i \phi_{ij} \bar{\xi}_j} |\xi\rangle,
\end{aligned} \tag{G34}$$

and carry out the Gaussian integral over $\bar{\xi}$ variables to arrive at (G33). Therefore, we can use the procedure explained in the previous appendix to compute the partial transpose $Z_{\mathcal{T}}$ with no difficulty. The result is shown in Fig. 4 of the main text. The important point here is that by using the Grassmann representation, the action of time-reversal operator becomes a change of variables in the space of Grassmann variables.

### 6. Fixed-point wave function

Here, we explicitly compute $Z_{\mathcal{T}}$ for the ground state wave function of the fixed-point Hamiltonian of Majorana chain introduced in Eq. (D3). Let $I_1 = \{1, 2, \ldots, M_1\}, I_2 = \{M_1 + 1, \ldots, M\}$ be adjacent intervals on closed circle $S$. First, we cut the system along the boundary of $I$. As a result, we are left with 6 Majorana zero modes

$$c_0^R \quad c_1^L \qquad\qquad c_{M_1}^R \quad c_{M_1+1}^L \qquad\qquad c_M^R \, c_{M+1}^L \tag{G35}$$

The Hilbert space of $I_1$, $I_2$, and $S \setminus I$ consist of the fermions

$$f_1^\dagger = \frac{c_1^L - i c_{M_1}^R}{2}, \qquad f_2^\dagger = \frac{c_{M_1+1}^L - i c_M^R}{2}, \qquad f_3^\dagger = \frac{c_{M+1}^L - i c_0^R}{2}, \tag{G36}$$

respectively. The corresponding terms in the Hamiltonian (Eq. (D3)) are

$$H_I = i(c_0^R c_1^L + c_{M_1}^R c_{M_1+1}^L + c_M^R c_{M+1}^L). \tag{G37}$$

The ground state of this Hamiltonian is given by

$$|\Psi\rangle = \frac{1}{2}[(1 + f_1^\dagger f_2^\dagger) f_3^\dagger |0\rangle + (f_1^\dagger + f_2^\dagger) |0\rangle]. \tag{G38}$$

The total density matrix for the ground state is

$$\rho = |\Psi\rangle\langle\Psi| = \frac{1}{4}\left[(1 + f_1^\dagger f_2^\dagger)|0\rangle\langle 0|(1 + f_2 f_1) + (f_1^\dagger + f_2^\dagger) f_3^\dagger |0\rangle\langle 0| f_3(f_1 + f_2) + \ldots\right], \tag{G39}$$

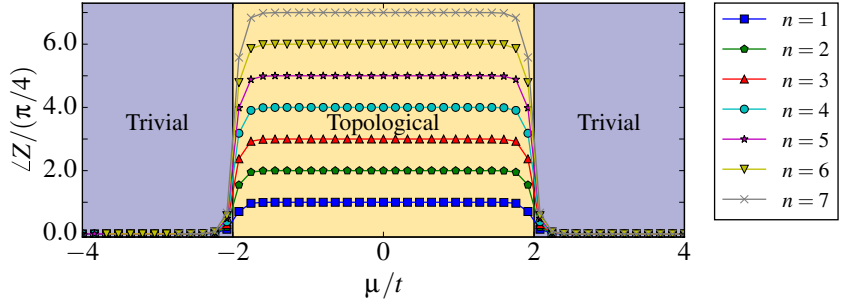

FIG. 7. Additivity of the complex phase for $Z_\mathcal{T}$ computed on a single Majorana chain with $n$ intervals. Here, $t = \Delta$, total system size is $N = 2nN_{\text{part}}$, and $N_{\text{part}} = 20$.

where ellipses refer to the off-diagonal terms. This gives rise to the reduced density matrix

$$\rho_I = \text{tr}_3 \rho = \frac{1}{4}\left[(1 + f_1^\dagger f_2^\dagger)|0\rangle\langle 0|(1 + f_2 f_1) + (f_1^\dagger + f_2^\dagger)|0\rangle\langle 0|(f_1 + f_2)\right]$$

$$= \frac{1}{4}\begin{pmatrix} 1 & 0 & 0 & 1 \\ 0 & 1 & 1 & 0 \\ 0 & 1 & 1 & 0 \\ 1 & 0 & 0 & 1 \end{pmatrix} = \frac{1}{4}(1 + \sigma_1 \otimes \sigma_1), \tag{G40}$$

where the density matrix is represented in the basis $\{|0\rangle, f_1^\dagger|0\rangle, f_2^\dagger|0\rangle, f_1^\dagger f_2^\dagger|0\rangle\}$ and $\sigma_i$ is Pauli matrices. This can also be written in terms of Majorana operators

$$\rho_I = \frac{1}{4}(1 + f_1^\dagger f_2^\dagger + f_2 f_1 + f_2^\dagger f_1 + f_1^\dagger f_2)$$

$$= \frac{1}{4}(1 - i c_{M_1}^R c_{M_1+1}^L). \tag{G41}$$

The partial transpose is given by

$$U_{I_1} \rho_I^{T_1} U_{I_1}^\dagger = \frac{1}{4}(1 + c_{M_1}^R c_{M_1+1}^L) = \frac{1}{4}(1 + i\sigma_1 \otimes \sigma_1), \tag{G42}$$

where $U_{I_1} = c_1^L$. The SPT invariant is found by

$$Z_\mathcal{T} = \text{tr}\left(\rho_I U_{I_1} \rho_I^{T_1} U_{I_1}^\dagger\right)$$

$$= \text{tr}\left[\frac{1}{4}(1 + \sigma_1 \otimes \sigma_1)\frac{1}{4}(1 + i\sigma_1 \otimes \sigma_1)\right]$$

$$= \frac{1+i}{4} = \frac{1}{2\sqrt{2}} e^{i\pi/4}. \tag{G43}$$

Similarly for $e^{i\theta} = -i$, we get $Z_\mathcal{T} = \frac{1}{2\sqrt{2}} e^{-i\pi/4}$. Thus, we reproduce the $\mathbb{Z}_8$ classification for the symmetry class BDI.

Let us now check that the $\mathbb{Z}_8$ topological phase derived from the partial transpose of the Majorana chain is additive under including additional fermion flavors or twist defects. In other words, if we include $n$ fermion flavors or introduce $n$ twist defects, the resulting complex phase computed by the partial transpose or the partial reflection will be $e^{in\pi/4}$. Let us first show this result analytically for the fixed-point wave function. Using Eq. (G41), the reduced density

matrix for $n$ flavors of fermions is given by

$$\rho_I = \bigotimes_{\alpha=1}^{n} \rho_{I,\alpha}$$
$$= \frac{1}{4^n} \prod_{\alpha=1}^{n} (1 - i c_{1,\alpha}^R c_{2,\alpha}^L). \tag{G44}$$

where $c_1$ and $c_2$ refer to the Majorana zero modes located at the ends of each interval and the index $\alpha = 1, \cdots, n$ is associated with fermion flavors corresponding to $n$ chains or $n$ intervals on a single chain. To obtain the partial transpose, we must first compute $\rho_I^{T_1}$,

$$\rho_I^{T_1} = \frac{1}{4^n} \prod_{\alpha=1}^{n} (1 - i\mathcal{R}(c_{1,\alpha}^R) c_{2,\alpha}^L)$$
$$= \frac{1}{4^n} \prod_{\alpha=1}^{n} (1 + c_{1,\alpha}^R c_{2,\alpha}^L). \tag{G45}$$

Therefore, the partial transpose $Z_{\mathcal{T}}$ is found by

$$Z_{\mathcal{T}} = \mathrm{tr}\,(\rho_I U_{I_1} \rho_I^{T_1} U_{I_1}^\dagger)$$
$$= \frac{1}{4^n} \mathrm{tr}\,[\prod_{\alpha=1}^{n}(1 - i c_{1,\alpha}^R c_{2,\alpha}^L) \prod_{\beta=1}^{n} c_{1,\beta}^L \prod_{\alpha'=1}^{n}(1 + c_{1,\alpha'}^R c_{2,\alpha'}^L) \prod_{\beta'=1}^{n} c_{1,\beta'}^L]$$
$$= \frac{1}{4^n} \mathrm{tr}\,[\prod_{\alpha=1}^{n}(1 + i) + \cdots]$$
$$= \frac{1}{(2\sqrt{2})^n} e^{in\pi/4}, \tag{G46}$$

where ellipses represent the off-diagonal terms.

For further verification of additivity, we numerically compute the partial transpose for $n$ separated intervals on a single chain. As we see in Fig. 7 the resulting complex phase in the topological regime is $e^{in\pi/4}$, as expected.

### Appendix H: Three-dimensional inversion symmetric topological superconductor (Class D + inversion)

As we have seen in the main text, the symmetry class D with inversion symmetry in three dimension is classified by $\mathbb{Z}_{16}$. To compute the partial inversion, we use the generating model in this symmetry class, which is given by the BdG Hamiltonian

$$\hat{H} = \frac{1}{2} \sum_{\mathbf{k}} \Psi^\dagger(\mathbf{k}) h(\mathbf{k}) \Psi(\mathbf{k}), \tag{H1}$$

on a cubic lattice, where

$$\Psi^\dagger(\mathbf{k}) = (f_\uparrow^\dagger(\mathbf{k}), f_\downarrow^\dagger(\mathbf{k}), f_\downarrow(-\mathbf{k}), -f_\uparrow(-\mathbf{k})), \tag{H2}$$

and

$$h(\mathbf{k}) = [-t(\cos k_x + \cos k_y + \cos k_z) - \mu]\,\tau_z + \Delta\,[\sin k_x \tau_x \sigma_x + \sin k_y \tau_x \sigma_y + \sin k_z \tau_x \sigma_z]. \tag{H3}$$

in which the $\tau$ and $\sigma$ matrices act on particle-hole and spin subspaces, respectively. This Hamiltonian also describes the $^3$He-B phase. The inversion symmetry in this model is defined by

$$\mathcal{I} f_\sigma(x,y,z) \mathcal{I}^{-1} = i f_\sigma(-x,-y,-z). \tag{H4}$$

We set $t = \Delta = 1$. This model exhibits three different topological phases as follows:

1. $1 < |\mu| < 3$: Topological I. This phase hosts a 2d gapless Majorana surface states. The corresponding interacting classification is $\mathbb{Z}_{16}$.

2. $|\mu| < 1$: Topological II. This phase supports an even number of 2d gapless Majorana surface states. It is topologically equivalent to a stack of 2d topological superconductors in the same symmetry class. Therefore, the corresponding classification is $\mathbb{Z}_8$.

3. $|\mu| > 3$: Trivial. No topological surface states.

We use the above definition of the inversion operator and transform a subset (cubic subsystem) of the lattice sites to compute the overlap function using Eq. (C12). The results are shown in Fig. 5 of the main text and completely match with those expected from the bordism theory.

### Appendix I: Partial reflection for projected wave functions

In this appendix, we present the results for the projected wave functions, i.e., we look at the wave function with a fixed number of particles and show that the $\mathbb{Z}_8$ classification survives in this case, too. The wave function is obtained by projecting the BCS wave function as in

$$|\Psi_{\bar{n}}\rangle = \mathcal{P}_{\bar{n}} |\Psi_{\text{BCS}}\rangle \tag{I1}$$

where $\mathcal{P}_{\bar{n}}$ is the projection operator to the averaged particle number of the BCS wave function, i.e.,

$$\bar{n} = \langle \Psi_{\text{BCS}} | \sum_x f_x^\dagger f_x | \Psi_{\text{BCS}} \rangle = \sum_k |v_k|^2 \tag{I2}$$

and

$$|\Psi_{\text{BCS}}\rangle = \prod_{k>0} (u_k + v_k f_k^\dagger f_{-k}^\dagger) |0\rangle . \tag{I3}$$

$|\Psi_{\bar{n}}\rangle$ is usually used as a good trial wave function with a fixed number of particles corresponding to the $|\Psi_{\text{BCS}}\rangle$. We construct the BCS wave function from the Majorana chain Hamiltonian (Eq. (1) of main text), where the coefficients are given by

$$u_k^2 = \frac{1}{2}\left(1 + \frac{\varepsilon_k - \mu}{\sqrt{(\varepsilon_k - \mu)^2 + \Delta_k^2}}\right), \tag{I4}$$

$$v_k^2 = \frac{1}{2}\left(1 - \frac{\varepsilon_k - \mu}{\sqrt{(\varepsilon_k - \mu)^2 + \Delta_k^2}}\right), \tag{I5}$$

in which $\varepsilon_k = -2t\cos k$ and $\Delta_k = 2\Delta \sin k$. To evaluate the partial reflection for the projected wave functions, we define a real-space representation of the wave function as introduced in Eq. (C15). Next, we use variational Monte Carlo to compute the wave function overlap

$$Z = \langle \Psi_{\bar{n}} | \tilde{\Psi}_{\bar{n}} \rangle = \frac{\sum_c \langle \Psi_{\bar{n}} | c \rangle \langle c | \tilde{\Psi}_{\bar{n}} \rangle}{\left[\sum_c |\langle c | \Psi_{\bar{n}} \rangle|^2 \sum_{c'} |\langle c' | \tilde{\Psi}_{\bar{n}} \rangle|^2\right]^{1/2}} \tag{I6}$$

where

$$|\tilde{\Psi}_{\bar{n}}\rangle = \mathcal{R}_{\text{part}} \mathcal{P}_{\bar{n}} |\Psi\rangle \tag{I7}$$

is the reflected wave function after the projection and the configuration $c = \{r_1, r_2, \ldots, r_{\bar{n}}\}$ refers to the positions of the $\bar{n}$ particles. The results are shown in Fig. 8 which are in excellent agreement with the mean-field BCS wave functions.

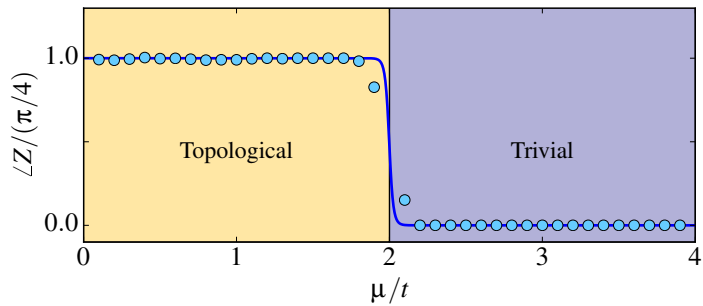

FIG. 8. Complex phase of the partial reflection for the projected BCS wave function (Eq. (I1)). Solid line shows the result for the mean-field BCS wave function (Fig. 3 in the main text). Here, we set $\Delta = t$, $N = 200$ and $N_{\text{part}} = 100$.

### Appendix J: Animations

We have provided four insightful animations to show the quantization and additivity of the complex phase of the non-local order parameters, partial reflection and partial transpose, introduced in the paper. Similar to the main text, we use the Kitaev Majorana chain in Eq. (1) for illustrations. In the following, we briefly explain the main features in each file. The system size details are shown in the animations. Although immaterial, we impose periodic boundary condition in all cases.

1. **Partial reflection**

    Here, we numerically compute the partial reflection (5) for the ground state of the Majorana chain (1) as a canonical model in the symmetry class D protected by the reflection symmetry.

    (a) `Quantization_of_partial_reflection.mp4`: This file shows the quantization of the complex phase associated with the partial reflection on a single chain. Clearly, the finite-size effects are more severe near the critical points (where the correlation length diverges). However, as we make the cross-cap larger, the complex phase becomes sharper and exhibits the quantization.

    (b) `Additivity_of_partial_reflection.mp4`: This file shows how the $\mathbb{Z}_8$ cyclic group emerges in our scheme for many-body SPT invariants. By introducing $n$ separate cross-caps over a long chain, the partial reflection yields plateaus at $Z = e^{i2\pi n/8}/(\sqrt{2})^n$ in the topological regime. This expectation value can be viewed either as $n$ copies of the Majorana chains or a space-time manifold generated by $n$ cross-caps (the cobordism group $\Omega_2^{\text{Pin}^-}$). From the former point of view, when $n = 8$ (i.e., eight copies of Majorana chain), the phase becomes $2\pi$ that is identical to 0 (i.e., the trivial phase).

2. **Partial transpose**

    Here, we numerically compute the partial transpose (6) for the ground state of the Majorana chain (1) as a canonical model in the symmetry class BDI protected by the time-reversal symmetry.

    (a) `Quantization_of_partial_transpose.mp4`: This file shows the quantization of the complex phase associated with the partial transpose. Similar to the partial reflection, the purpose of this animation is to illustrate the fact that the complex phase is quantized as long as the size of the cross-cap is sufficiently large compared to the correlation length.

    (b) `Additivity_of_partial_transpose.mp4`: This file shows how the $\mathbb{Z}_8$ cyclic group emerges in the partial transpose. By creating $n$ separate cross-caps in the system, the partial transpose gives plateaus at $Z = e^{i2\pi n/8}/(2\sqrt{2})^n$ in the topological regime. The trend clearly shows that when $n = 8$ (i.e., eight copies of Majorana chain), the phase becomes $2\pi$ that is identical to 0 (i.e., the trivial phase).

---



\end{document}